\documentclass[12pt,a4paper]{article}
\usepackage{amsmath}
\usepackage{amssymb}
\def\datta{Nilanjana Datta\footnote{email: n.datta@statslab.cam.ac.uk.}  \&  Tony Dorlas\footnote{email:dorlas@stp.dias.ie; on leave from the
Dublin Institute of Advanced Studies, School of Theoretical Physics, 10 Burlington Road, 
Dublin 4, Ireland}\\ {\slshape Statistical Laboratory} \\
{\slshape Centre for Mathematical Sciences} \\ {\slshape
University of Cambridge} \\ {\slshape Wilberforce Road, Cambridge
CB3 0WB, U.K.} \\}

\def\one{{\mbox{\bf 1}}}
\def\disp{\displaystyle}

\newcommand{\half}{\frac{1}{2}}
\newcommand{\third}{\frac{1}{3}}
\newcommand{\Tr}{\mbox{\rm Tr}\,}
\newcommand{\Trace}{\mbox{\rm Tr}\,}
\newcommand{\tr}{\mbox{\rm Tr}\,}
\newcommand{\non}{\nonumber}
\newcommand{\ul}{\underline}
\newcommand{\opplus}{\bigoplus}
\newcommand{\ol}{\overline}
\newcommand{\uj}{\ul{j}}
\newcommand{\iden}{\bf{1}}

\newtheorem{theorem}{Theorem}
\newtheorem{lemma}{Lemma}

\newtheorem{corollary}{Corollary}

\def\be{\begin{equation}}
\def\ee{\end{equation}}
\def\bea{\begin{eqnarray}}
\def\eea{\end{eqnarray}}
\def \outlineby #1#2#3{\vbox{\hrule\hbox{\vrule\kern #1%
\vbox{\kern #2 #3\kern #2}\kern #1\vrule}\hrule}}%
\def \endbox {\outlineby{4pt}{4pt}{}}%
\newcommand{\qed}{\hfill \endbox}

\def\reff#1{(\ref{#1})}

\begin{document}

\title{Classical capacity of quantum channels with general
Markovian correlated noise}
\author{\datta}

\maketitle

\begin{abstract} The classical capacity of a quantum channel with {\em{arbitrary}}
Markovian correlated noise is evaluated. For the general case of a channel with long-term
memory, which corresponds to a Markov chain which does not converge to equilibrium, 
the capacity is expressed in terms of the communicating classes of the Markov chain. 
For an irreducible and aperiodic Markov
chain, the channel is forgetful, and one retrieves the known expression
\cite{KW} for the capacity.
\end{abstract}

\newpage

\section{Introduction}

Shannon, in his celebrated Noisy Channel Coding Theorem \cite{shannon},
obtained an explicit expression for the channel capacity
of discrete, memoryless\footnote{
For such a channel, the noise affecting successive input states,
is assumed to be perfectly uncorrelated.},
classical channels. The first
rigorous proof of this fundamental theorem was provided by
Feinstein \cite{Feinstein}. He used a packing argument (see e.g.\cite{Khinchin}
) to find a
lower bound to the maximal number of codewords that can be sent
through the channel reliably, i.e., with an arbitrarily
low probability of error.
More precisely, he proved that for any given $\delta >0$, and sufficiently
large number, $n$, of uses of a memoryless classical channel, the lower bound
to the maximal number, ${N_n}$, of codewords that can be transmitted through the
channel reliably, is given by $${N_n} \ge 2^{n(H(X:Y)-\delta)}.$$ Here
$H(X:Y)$ is the mutual information of the random variables $X$ and $Y$,
corresponding to the input and the output of the channel, respectively.
This lower bound implies that for $n$ large enough, any real number 
$R <C = \max \, H(X:Y)$,
(the maximum being taken over all possible input distributions), 
at least $N_n = [2^{nR}]$
classical messages can be transmitted through the channel reliably. In
other words, any rate $R <C$ is {\em{achievable}}.

The {assumption} that 
noise is {uncorrelated} between {successive uses} of a channel 
is not realistic. Hence {memory effects }need to be taken into account.
In this paper we consider the transmission of classical
information through a class of quantum channels with memory.
The {first model} of such a channel was studied by
{Macchiavello and Palma} \cite{chiara}. They showed that the
transmission of {classical information} through {two successive
uses} of a {quantum depolarising channel, with Markovian
correlated noise,} is enhanced by using inputs entangled over the
two uses. A more general model of a quantum channel with 
memory was introduced by Bowen and Mancini
\cite{bowen} and also studied by Kretschmann and Werner \cite{KW}.
In particular, in \cite{KW}, the capacities of
a class of quantum channels with memory, the so-called {\em{forgetful channels}}
were evaluated. Similar results were obtained by Bjelakovi\'c and Boche \cite{berlin}.
Further, in \cite{DD1}, the classical capacity of a class of quantum channels with 
long-term memory was obtained. The memory of the channel considered in \cite{DD1}
can be viewed as a special case of a general Markovian memory, where the 
Markov chain is aperiodic but not irreducible, and hence does not 
converge to equilibrium. Recently, there was a generalization of the result of \cite{DD1}
by Bjelakovi\'c and Boche, who in \cite{berlin2} obtained the classical capacities
of compound and averaged quantum channels.

Another interesting special case of a channel with long-term memory is that
in which the memory is described by a periodic Markov chain. A simple example
of this is a channel given by alternating applications of two completely positive
trace preserving (CPT) maps $\Phi_1$ and $\Phi_2$, with the 
first map being $\Phi_1$ or $\Phi_2$ with probability $1/2$.

In this paper we study channels with {\em{arbitrary}} Markovian correlated 
noise. This includes, in particular, the above special cases. We show that the
capacity in the general case can be expressed in terms of the communicating
classes of the underlying Markov chain.

We start the main body of our paper with some preliminaries in
Section \ref{prelim}. In Section \ref{main}, the quantum channel is defined
and its capacity is stated in the main theorem, Theorem \ref{main1},
of this paper. In Section \ref{ergodic}, we prove a special case of the
direct part of this theorem, corresponding to a Markov chain which
converges to equilibrium and is hence forgetful. This section therefore provides
an alternative proof of the result of Kretschmann and Werner \cite{KW} for the
classical capacity of such a channel. This proof is extended to the case of
an arbitrary Markov chain in Section \ref{general}. In the latter, we
employ the idea of adding a preamble to the codewords (as was done in \cite{DD1})
in order to distinguish between the different communicating classes of the Markov
chain. The proof of the (weak) converse part of our main result (Theorem \ref{main1})
is given in Section \ref{converse}.

\section{Mathematical Preliminaries}
\label{prelim}
Let $\cal H$ and $\cal K$ be given finite-dimensional Hilbert
spaces and denote by ${\cal B}({\cal H})$ the algebra of linear
operators on $\cal H$. We also consider the tensor product
algebras ${\cal A}_n = {\cal B}({\cal H}^{\otimes n})$ and the
infinite tensor product C$^*$-algebra obtained as the strong
closure
\begin{equation} {\cal A}_\infty = \overline{ \bigcup_{n=1}^\infty
{\cal A}_n}, \end{equation} where we embed ${\cal A}_n$ into
${\cal A}_{n+1}$ in the obvious way. Similarly, we define ${\cal
B}_n = {\cal B}({\cal K}^{\otimes n})$ and ${\cal B}_\infty$. 
A {\em{state}} on an algebra ${\cal{A}}$ is a positive linear functional
$\phi$ on ${\cal{A}}$ with $\phi({\iden})= 1$, where $\iden$ denotes 
identity operator. If ${\cal{A}}$ is finite-dimensional then there 
exists a density matrix $\rho_\phi$ (i.e., a positive operator
with ${\tr}\rho_\phi=1$) such that $\phi(A) = {\tr}(\rho_\phi A),$
for any $A \in {\cal{A}}$. We
denote the states on ${\cal A}_\infty$ by ${\cal S}({\cal
A}_\infty)$, those on ${\cal A}_n$ by ${\cal S}({\cal
A}_n)$,etc.

\section{A quantum channel with classical memory}
\label{main}
Let there be given a Markov chain on a finite state space $I$
with transition probabilities $\{q_{ii'}\}_{i,i' \in I}$ and let
$\{\gamma_i\}_{i \in I}$ be an invariant distribution for this
chain, i.e.
\begin{equation} \gamma_{i'} = \sum_{i \in I} \gamma_{i} q_{ii'}.
\label{eqdist}
\end{equation}
Moreover, let $\Phi_i: {\cal B}({\cal H}) \to {\cal B}({\cal K})$
be given completely positive trace-preserving (CPT) maps for each
$i \in I$. Then we define a quantum channel with Markovian correlated noise,
by the CPT map
$\Phi_\infty: {\cal S}({\cal A}_\infty) \to {\cal S}({\cal
B}_\infty)$ on the states of ${\cal A}_\infty$ by
\begin{equation} (\Phi_\infty) (\phi) (A) = \sum_{i_1,\dots,i_n
\in I} \gamma_{i_1} q_{i_1i_2} \dots q_{i_{n-1}i_n} \Tr \left[
(\Phi_{i_1} \otimes \dots \otimes \Phi_{i_n})(\rho_{\phi_n}) \, A
\right] \label{mem}
\end{equation} for $A \in {\cal B}_n$. Here, $\phi_n$ is the
restriction of $\phi$ to ${\cal A}_n$ and $\rho_{\phi_n}$ its
density matrix. It is easily seen, using the property
(\ref{eqdist}), that this definition is consistent and defines a
CPT map on the states of ${\cal A}_\infty$, and moreover, that it
is translation-invariant (stationary).

We denote the transpose action of
the restriction of $\Phi_\infty$ to ${\cal S}({\cal
A}_n)$ by $\Phi^{(n)}: {\cal B}({\cal H}^{\otimes n}) \to {\cal
B}({\cal K}^{\otimes n})$, i.e., 
$${\tr}\bigl(\Phi^{(n)}(\rho_\phi)A \bigr) =(\Phi_\infty(\phi))(A),$$
for a density matrix $\rho_\phi \in {\cal{B}}({\cal{H}}^{\otimes n})$,
$\phi \in  {\cal S}({\cal A}_n)$.

Note that
\begin{equation}
\Phi^{(n)}(\rho^{(n)}) = \sum_{i_1,\dots,i_n \in I} \gamma_{i_1}
q_{i_1i_2} \dots q_{i_{n-1}i_n} (\Phi_{i_1} \otimes \dots
\otimes \Phi_{i_n})(\rho^{(n)}). \label{mem2}\end{equation}

Let us consider the transmission of classical information through
$\Phi^{(n)}$. Suppose Alice has a set of
messages, labelled by the elements of the set ${\cal{M}}_n = \{1,2,
\ldots, M_n\},$ which she would like to communicate to Bob, using
the quantum channel $\Phi$. To do this, she encodes each message
into a quantum state of a physical system with Hilbert space
${\cal{H}}^{\otimes n}$, which she then sends to Bob through
$n$ uses of the quantum channel. In order to infer the message
that Alice communicated to him, Bob makes a measurement (described
by POVM elements) on the state that he receives. The encoding and
decoding operations, employed to achieve reliable transmission of
information through the channel, together define a quantum error
correcting code (QECC). More precisely, a code ${\cal{C}}^{(n)}$
of size $N_n$ is given by a sequence $\{\rho_i^{(n)},
E_i^{(n)}\}_{i=1}^{N_n}$ where each $\rho_i^{(n)}$ is a state in
${\cal{B}}({\cal{H}}^{\otimes n})$ and each $E_i^{(n)}$ is a
positive operator acting in ${\cal{K}}^{\otimes n}$, such that
$\sum_{i=1}^{N_n} E_i^{(n)} \le {\one}^{(n)}$.
Here, ${\one}^{(n)}$ denotes the identity operator in
${\cal{B}}({\cal K}^{\otimes n})$.
Defining
$E_0^{(n)} = {\one}^{(n)} - \sum_{i=1}^{N_n} E_i^{(n)}$, yields a
Positive Operator-Valued Measure (POVM) $\{E_i^{(n)}\}_{i=0}^{N_n}$
in ${\cal{K}}^{\otimes n}$. An output $i\ge 1$
would lead to the inference that the state (or codeword)
$\rho_i^{(n)}$ was transmitted through the channel $\Phi^{(n)}$,
whereas the output $0$ is interpreted as a failure of any
inference. 
The average probability of error for the code ${\cal{C}}^{(n)}$ is given by
\begin{equation}
P_e({\cal{C}}^{(n)}):= \frac{1}{N_n} \sum_{i=1}^{N_n} \left(1 - {\Trace}\bigl(
\Phi^{(n)}(\rho_i^{(n)}) E_i^{(n)}\bigr)\right),
\label{codeerr}
\end{equation}
If there exists an
$n_0 \in {\mathbb{N}}$ such that for all $n \ge n_0$, there exists a sequence
of codes $\{{\cal{C}}^{(n)}\}_{n=1}^\infty$, of sizes $N_n \ge 2^{nR}$,
for which $P_e({\cal{C}}^{(n)})\rightarrow 0$ as $n \rightarrow \infty$, then
$R$ is said to be an {\em{achievable}} rate.
\medskip

The {\em{classical capacity}} of ${\Phi}$ is defined as
\begin{equation}
C(\Phi) := \sup R,
\end{equation}
where $R$ is an achievable rate.

Let ${\cal C}$ be the
set of communicating classes, $C$, of the Markov chain \cite{Norris}
for which
\begin{equation} \gamma_C = \sum_{i \in C} \gamma_i > 0.
\end{equation} Any other classes can be disregarded. 
For $C \in {\cal C}$ we define
\begin{equation} \Phi^{(n)}_C(\rho^{(n)}) := \frac{1}{\gamma_C}
\sum_{i_1,\dots,i_n \in C} \gamma_{i_1} q_{i_1i_2} \dots
q_{i_{n-1}i_n} (\Phi_{i_1} \otimes \dots \otimes
\Phi_{i_n})(\rho^{(n)}), \end{equation}
which represents the restriction of the classical memory of
the channel to the class $C$. Notice that the Markov chain
restricted to $C \in {\cal C}$ is necessarily irreducible,
and is either aperiodic or periodic with a single period. In fact,
$${\cal C} = {\cal C}_{aper} \cup {\cal C}_{per},$$
where ${\cal C}_{aper}$ denotes the set of communicating classes
in ${\cal C}$ which are aperiodic, while ${\cal C}_{per}$ denotes the set of 
communicating classes in ${\cal C}$ which are aperiodic.

If $C \in {\cal C}_{aper}$, we define, for any ensemble $\{p_j^{(n)}, \rho_j^{(n)}\}$ of states on 
${\cal H}^{\otimes n}$, the {\em{mean Holevo quantity}} for the
class $C$ as
\begin{equation}
{\bar{\chi}}_C^{(n)}(\{p_j^{(n)},\rho_j^{(n)}\}) := \frac{1}{n} \left[ S \left(
\sum_{j} p_{j}^{(n)}
\Phi_C^{(n)}(\rho_{j}^{(n)}) \right) - \sum_{j} p_{j}^{(n)} S \left(\Phi^{(n)}_C
(\rho_{j}^{(n)}) \right) \right]. \label{39}\end{equation} 
If $C \in {\cal C}_{per}$ is periodic, with period $L$, then
$C=\{i_0,i_1,\dots,i_{L-1}\}$ for certain $i_0,\dots,i_{L-1} \in I$, and
$q_{i_{k}i_{k+1}} = 1$ for $k=0,\dots,L-2$ and $q_{i_{L-1}i_0}=1$. In
this case, \begin{equation} \gamma_i = \frac{1}{L} \gamma_C
\quad (i \in C) \end{equation} and we set 
\begin{equation}
{\bar{\chi}}_C^{(n)}
(\{p_j^{(n)},\rho_j^{(n)}\}) = \frac{1}{nL} \sum_{i \in C}
\chi_{C,i}^{(n)}(\{p_j^{(n)},\rho_j^{(n)}\}), \label{311} \end{equation} 
where for $k \in \{0,1,\ldots, L-1\}$, 
\begin{equation}
\chi_{C,i_k}^{(n)}(\{p_j^{(n)},\rho_j\}) = S \left( \Phi_{i_k} \otimes \Phi_{i_{k+1}} \dots 
\otimes \Phi_{i_{k+n-1}}({\bar \rho}^{(n)}) \right) -
{\bar S}_{i_k}^{(n)}, \end{equation}
(the indices in the subscripts being taken modulo $L$), 
with 
\begin{equation} {\bar \rho}^{(n)} =
\sum_j p_j^{(n)} \rho_j^{(n)}, \mbox{ and } {\bar S}_{i_k}^{(n)} = \sum_j p_j^{(n)}
S(\Phi_{i_k} \otimes \Phi_{i_{k+1}} \dots \otimes \Phi_{i_{k+n-1}} (\rho_j^{(n)})). 
\label{313}
\end{equation} 

Our main result is the following theorem. We use the standard notation 
$\wedge$ for minimum and $\vee$ for maximum.
\begin{theorem}
\label{main1}
The classical capacity of a quantum channel with arbitrary 
Markovian correlated noise, defined by \reff{mem},
is given by 
\begin{equation}
C(\Phi) = \lim_{n \to \infty}
\sup_{\{p_j^{(n)},\rho_j^{(n)}\}} \left[ \bigwedge_{C \in {\cal C}}
{\bar{\chi}}_C^{(n)} (\{p_j^{(n)},\rho_j^{(n)}\}) \right]. \label{314}
\end{equation}
%The existence of this limit is proved in an identical fashion to Lemma~\ref{lemma1}.
\end{theorem}
The existence of the limit in \reff{314} is proved in Lemma \ref{lema1} of Appendix A.

Before proving Theorem \ref{main1}, we consider the special case in which 
the Markov chain has a single communicating class, and the latter is
aperiodic and irreducible.

\section{Ergodic memory case}
\label{ergodic}
\setcounter{equation}{0}

In this section we assume that the underlying Markov chain is aperiodic and irreducible 
(see e.g. \cite{Norris}) so that in particular, 
the invariant distribution, $\{\gamma_i\}_{i \in I}$, is unique. It is well-known that the corresponding 
Markov chain is ergodic and consequently the output states of the 
channel are also ergodic. In this case, the Markov chain satisfies
the property of {\em{convergence to equilibrium}}, i.e., 
$$p_{ij}^{(n)} \rightarrow \gamma_j \quad {\hbox{as}} \,\, n \rightarrow \infty,$$
where $p_{ij}^{(n)}$ denotes the $n$-step transition probability from the state
$i$ to the state $j$, $(i,j\in I)$.
This implies that the correlation in the noise,
acting on successive inputs to the channel, dies out after a sufficiently
large number of uses of the channel. Hence, in this case the channel belongs
to the class of channels introduced and studied by Kretschmann and Werner \cite{KW},
and referred to as {\em{forgetful channels}}.

Suppose that $\{p_j^{(n)},\rho_j^{(n)}\}_{j=1}^{J(n)}$ is
a sequence of states given by density matrices $\rho_j^{(n)}$ on ${\cal H}^{\otimes n}$ with
probabilities $p_j^{(n)}$, $\sum_{j=1}^{J(n)} p_j^{(n)} = 1$. 

The Holevo quantity for the channel restricted to ${\cal
A}_n$ is given by \begin{equation}
\chi(\{p_j^{(n)},\Phi^{(n)}(\rho_j^{(n)})\}) = S \left(
\sum_{j=1}^{J(n)} p_j^{(n)} \Phi^{(n)}(\rho_j^{(n)}) \right) -
\sum_{j=1}^{J(n)} p_j^{(n)} S(\Phi^{(n)}(\rho_j^{(n)})) 
\ee
The classical capacity of a quantum channel with classical ergodic memory 
is stated in the following theorem, which is a special case of Theorem \ref{main1}.
\begin{theorem}
\label{main2}
The classical capacity of a quantum channel with memory defined by \reff{mem},
where the underlying Markov chain is aperiodic and irreducible, is 
given by 
\begin{equation}
{\chi}^*(\Phi) = \lim_{n \to \infty} \frac{1}{n}
\sup_{\{p_j^{(n)},\rho_j^{(n)}\}} \chi(\{p_j^{(n)},\Phi^{(n)}
(\rho_j^{(n)})\}) \label{cap1}
\ee
\end{theorem}
The existence of the limit in \reff{cap1} is proved in Lemma \ref{lema1} of Appendix A.

This expression for the capacity was in fact stated and proved in \cite{KW}. We present
an alternative proof which can then be extended to the case of a general 
Markov chain. The latter is done in Section \ref{general}.

The direct part of Theorem \ref{main2}, i.e., the achievability of any rate
$R < \chi^*(\Phi)$, follows from Lemma \ref{feinmem} given below, which is 
itself a generalization of the Quantum Feinstein Lemma for a memoryless
channel \cite{DD0, DD1}. The weak converse part of Theorem \ref{main2}
is proved in the general case in Section \ref{converse}.

\subsection{Quantum version of Feinstein's Lemma}

\begin{lemma}
\label{feinmem} Let $\Phi_\infty$ denote a quantum memory channel
with Markovian correlated noise, defined by (\ref{mem}). Suppose
that the Markov chain is aperiodic and irreducible. Let ${\chi}^*
= {\chi}^*(\Phi)$
be given by (\ref{cap1}). Given $\epsilon> 0$, there exists $n_0
\in {\mathbb{N}}$ such that for all $n\geq n_0$ there exist at least $N \geq
2^{n({{\chi}^*}-\epsilon)}$ states with density matrices
${\tilde \rho}^{(n)}_1, \dots, {\tilde \rho}^{(n)}_N \in {\cal
B}({\cal H}^{\otimes n})$, and positive operators $E^{(n)}_1,\dots,
E^{(n)}_N \in {\cal B}({\cal K}^{\otimes n})$ such that
$\sum_{k=1}^N E^{(n)}_k \leq \one^{(n)}$ and
\begin{equation} \tr  \left[ \Phi^{(n)} \left( {\tilde
\rho}^{(n)}_k \right) E^{(n)}_k \right]  > 1-\epsilon.
\end{equation}
\end{lemma}
\textbf{Proof.} Choose $l_0$ so large that \begin{equation} \left|
\frac{1}{l_0} \sup_{\{p_j^{(l_0)},\rho_j^{(l_0)}\}}
\chi(\{p_j^{(l_0)},\Phi^{(l_0)} (\rho_j^{(l_0)})\}) - \chi^*
\right| \leq \frac{1}{6} \epsilon. \label{chiapprox} \end{equation}
Then assume that the supremum is attained for an ensemble
$\{p_j^{(l_0)}, \rho_j^{(l_0)} \}_{j=1}^{J}$, for a finite $J$.

Denote for $m \in {\mathbb{N}}$
\begin{equation} {\bar \sigma}_{ml_0} = \Phi^{(m l_0)}
\left( {\bar \rho}_{l_0}^{\otimes m} \right), \label{sigmabar}
\end{equation} where \begin{equation} {\bar \rho}_{l_0} =
\sum_{j=1}^J p_j^{(l_0)} \rho_j^{(l_0)}. \end{equation} These
states form a compatible system of states on $\{{\cal B}_{m l_0}\}_{m=1}^\infty$ and
hence a state ${\bar\phi}_\infty$ on ${\cal B}_\infty$ by
\begin{equation} 
{\bar \phi}_\infty (A) = \Tr\, ({\bar\sigma}^{ml_0} A) 
\label{barphi}
\end{equation} if $A \in {\cal B}_{m l_0}$. This
state is clearly $l_0$-periodic, i.e. invariant under translations
over multiples of $l_0$ . Therefore, the mean entropy
\begin{equation} S_M({\bar\phi}_\infty) := \lim_{m \to \infty}
\frac{1}{m} S \left({\bar\sigma}^{ml_0} \right) = \inf_{m \in {\mathbb{N}}}
\frac{1}{m} S \left({\bar\sigma}^{ml_0} \right)
\end{equation} exists.

For $l_0$ sufficiently large, the mean entropy  $S_M({\bar\phi}_\infty)$ 
is close to $S \left({\bar\sigma}_{l_0}\right)$, the von Neumann entropy
of the average output of $l_0$ uses of the channel.
This is stated in the following lemma.

\begin{lemma} \label{lem24} 
Given $\epsilon > 0$ there exists $L > 0$ such that for
$l_0 \geq L$, \begin{equation} \left| \frac{1}{l_0} S_M({\bar
\phi}_\infty) - \frac{1}{l_0} S(\Phi^{(l_0)}({\bar \rho}_{l_0})) \right|
< \frac{\epsilon}{8} 
\end{equation} 
\label{lemma1} \end{lemma}
Here ${\bar \phi}_\infty$ is given by \reff{barphi}.
The proof is similar to that of Lemma~\ref{lemma1}.

Henceforth $l_0$ is fixed to a value such that Lemma \ref{lem24} and
\reff{chiapprox} hold. For notational simplicity, explicit
dependence on $l_0$ is often suppressed.

The proof of Lemma \ref{feinmem} requires the sequence of lemmas given
below. 

\begin{lemma} The state ${\bar\phi}_\infty$ is strongly
clustering and hence completely ergodic for $l_0$-shifts, i.e.,
for any $A,B \in {\cal B}_{ml_0}$,
\begin{equation} \lim_{k \to \infty} {\bar\phi}_\infty
(A\,\tau^{kl_0} (B)) = \Tr({\bar\sigma}_{ml_0}\,A) \,
\Tr({\bar\sigma}_{ml_0}\,B).
\end{equation}
\label{lemma2}
\end{lemma}
{\bf{Proof}}
The proof is standard and relies on the fact that the expectations
of $A$ and $B$ in the state ${\bar\phi}_\infty$ decouple as their 
supports are separated by a sufficiently large distance. This is 
because 
\begin{equation}
\sum_{i_{2},i_3, \ldots, i_k} q_{i_1i_2} \dots q_{i_{k-1}i_k}
g(i_k)  \to \sum_i \gamma_i g(i), \label{ergchain} \end{equation}
as $k \to \infty$, {{for any function}} $g(i)$, since the
Markov chain is irreducible and aperiodic.
\qed

In the following we denote ${\cal K}^{\otimes l_0}$ by ${\cal K}_{l_0}$. We
also use the following lemma, which is proved in Appendix B.

\begin{lemma} \label{lem44}
For any $\delta >0$, there exists $m_1 \in {\mathbb{N}}$ such that for all $m \geq m_1$ there
exists a subspace ${\cal T}_{\epsilon}^{(m)} \subset {\cal
K}_{l_0}^{\otimes m}$ with projection ${\bar P}_{ml_0}$ such that
\begin{equation} {\bar P}_{ml_0}\, {\bar \sigma}_{ml_0} \,{\bar P}_{ml_0} \leq
2^{-m [S_M({\bar\phi}_\infty ) - \frac{1}{4} \epsilon]} \one^{(ml_0)} \label{bartyp}
\end{equation} and
\begin{equation} \Tr \left( {\bar\sigma}_{ml_0}\,{\bar P}_{ml_0} \right) >
1-\delta^2.
\end{equation}
Here $\one^{(ml_0)}$ denotes the identity operator in ${\cal{B}}({\cal{K}}^{\otimes ml_0})$.
\end{lemma}

In order to obtain the first term in the expression \reff{cap1} for the capacity, we need
to be able to replace $S_M({\bar{\phi}}_\infty)$ in the above lemma by
$S({\bar \sigma}_{l_0})$. This is possible due to Lemma \ref{lem24}.

We need an analogous result to Lemma~\ref{lem44} for the second
term in the expression \reff{cap1} of ${\chi}^*(\Phi)$. This is stated
in Lemma \ref{lem46} (which is proved in Appendix C). It 
uses Lemma \ref{lem45}, given below. To formulate these lemmas, 
we define density matrices $\Sigma_{ml_0}$ in algebras $$ {\cal M}_{ml_0} =
\bigoplus_{j_1,\dots,j_m = 1}^J 
{\cal B} ({\cal K}_{l_0}^{\otimes m}) $$ by
\begin{equation} \Sigma_{ml_0} = \bigoplus_{j_1,\dots, j_m} p_{\ul{j}}^{(m)}
 \Phi^{(ml_0)} \left(
\rho_{j_1}^{(l_0)} \otimes \dots \otimes \rho_{j_m}^{(l_0)} \right),
\end{equation}
where $ p_{\ul{j}}^{(m)} = \prod_{\alpha=1}^m p_{j_\alpha}^{(l_0)}$ and $\rho_{j}^{(l_0)}$, 
$j\in \{1,2,\ldots,J\}$, belongs to the maximising ensemble (c.f. \reff{chiapprox}). In the
following we denote $ \rho_{\ul{j}}^{(ml_0)} = \rho_{j_1}^{(l_0)} \otimes \dots \otimes \rho_{j_m}^{(l_0)}$,
with $\ul{j} = (j_1,j_2,\ldots, j_m)$, for any $m \in {\mathbb{N}}$. 

\begin{lemma} \label{lem45}
There exists a unique translation-invariant state $\psi_\infty$ on
${\cal M}_\infty = \ol{\bigcup_{m=1}^\infty {\cal M}_{ml_0}}$ such that
\begin{equation} \psi_\infty (A) = \Tr \left( \Sigma_{ml_0}\,A \right)
\end{equation} for $A \in {\cal M}_{ml_0}$. Moreover, this state is
strongly clustering and therefore completely ergodic.
\end{lemma}
{\bf{Proof.}} The proof of this lemma is similar to that of
Lemma~\ref{lemma2}.\qed

Note that the mean entropy of $\psi_\infty$ is given by 
\begin{equation} S_M(\psi_\infty) = \lim_{k \to \infty} \frac{1}{k} S( \Sigma_{kl_0}), \end{equation} 
where \begin{eqnarray} S(\Sigma_{kl_0}) &=& \sum_{j_1,\dots,j_k} S \left( p_{\ul{j}}^{(k)} 
\Phi^{(kl_0)}(\rho_{j_1}^{(l_0)} \otimes \dots \otimes \rho_{j_k}^{(l_0)}) \right) \non \\ 
&=& \sum_{\ul{j}} p_{\ul{j}}^{(k)} S\left(
\Phi^{(kl_0)}(\rho_{\ul{j}}^{(kl_0)}) \right) + k\,H \left( \{p_j^{(l_0)}\}_{j=1}^J \right). 
\end{eqnarray}
We define
\begin{equation} {\bar S}_M\equiv {\bar S}_M\left( \{p_j^{(l_0)}, \rho_j^{(l_0)}\} \right) 
:= \lim_{k \rightarrow \infty} \frac{1}{k} \sum_{\ul{j}} p_{\ul{j}}^{(k)} S\left(
\Phi^{(kl_0)}(\rho_{\ul{j}}^{(kl_0)}) \right). \end{equation} 

\begin{lemma} \label{lem46}
Given $\delta > 0$, there exists $m_2 \in {\mathbb{N}}$ such that for
all $m \geq m_2$ there exist, for all $\ul{j} = (j_1,\dots,j_m)
\in \{1,\dots,J\}^m$, one-dimensional subspaces ${\cal
T}_{\ul{j},\ul{k}}^{(m)}$  of ${\cal K}_{l_0}^{\otimes m}$\
(indexed by $\ul{k}$ in some set $ T_{\ul{j},\epsilon}^{(m)}$) with
projections $\pi^{(ml_0)}_{\ul{j},\ul{k}}$ in the $\ul{j}$-th
component of ${\cal M}_{ml_0}$, such that for all $\ul{k} \in
T_{\ul{j},\epsilon}^{(m)}$,
\begin{equation} \left| \frac{1}{m} \log \omega_{\ul{j},\ul{k}}^{(ml_0)} +
{\bar S}_M (\{p_j^{(l_0)},\rho_j^{(l_0)}\}) \right| < \frac{\epsilon}{4},
\label{typ_j}
\end{equation} where $\omega_{\ul{j},\ul{k}}^{(ml_0)} = \Tr \left(
\Phi^{(ml_0)}(\rho_{\ul{j}}^{(ml_0)}) \pi_{\ul{j},\ul{k}}^{(ml_0)}
\right)$, and
\begin{equation} \psi_\infty \left( \bigoplus_{\ul{j}}
\bigoplus_{\ul{k} \in T_{\ul{j},\epsilon}^{(m)}}
\pi_{\ul{j},\ul{k}}^{(ml_0)} \right) > 1-\delta^2.
\end{equation}
\end{lemma}

\noindent
{\bf{Proof}} See Appendix C.
%-------------------------------------------------------------------

We now continue the proof of the theorem. In the following we
denote  \begin{equation} P_{\ul{j}}^{(ml_0)} = \bigoplus_{\ul{k} \in
T_{\ul{j},\epsilon}^{(m)}} \pi_{\ul{j}, \ul{k}}^{(ml_0)} \end{equation}
%and we put $\sigma_{\ul{j}}^{(n)} =
%\Phi^{(n)}(\rho_{\ul{j}}^{(n)})$.
The remainder of the proof is in fact analogous to that for the
case of a memoryless channel (see \cite{DD0}, \cite{DD1}), 
so we only sketch the main steps.

For arbitrary $n$, let $m= [n/l_0]$ and denote, 
${\bar{\Pi}}_n = {\bar P}_{ml_0} \otimes \one_{n-ml_0}$, $
{\Pi}_{\ul{j}}^{(n)} = P_{\ul{j}}^{(ml_0)} \otimes \one_{n-ml_0}$ and
${\bar{\sigma}}^{(n)} = \tr_{(m+1)l_0 -n} {\bar{\sigma}}_{(m+1)l_0}$ . Now
let $N=N(n)$ be the maximal number of
states ${\tilde \rho}_1^{(n)}, \dots, {\tilde \rho}_N^{(n)}$ on
${\cal H}^{\otimes n}$ for which there exist positive operators 
$E_1^{(n)},\dots, E_N^{(n)}$ on ${\cal K}^{\otimes n}$,
of the form $E_j^{(n)} = {\cal{E}}_j^{ml_0} \otimes {\iden}_{n-ml_0}$, such that
\begin{enumerate} \item[(i)] $ \sum_{k=1}^N E_k^{(n)} \leq {\bar{\Pi}}_n$ and
\item[(ii)]  $\tr [\,{\tilde \sigma}_k^{(n)} E_k^{(n)} ] >
1-\epsilon$ and \item[(iii)] $\tr [\,{\bar \sigma}^{(n)} E_k^{(n)}
] \leq 2^{-n[{\chi}^*(\Phi) -\frac{2}{3} \epsilon]}$.
\end{enumerate}
Here ${\tilde \sigma}_k^{(n)} = \Phi^{(n)}({\tilde
\rho}_k^{(n)})$.

For any given $\ul{j} \in \{1,\dots,J\}^m$ define
\begin{equation} V_{\ul{j}}^{(n)} = \left( {\bar{\Pi}}_n - \sum_{k=1}^N
E_k^{(n)} \right)^{1/2} {\bar{\Pi}}_n {\Pi}_{\ul{j}}^{(n)} {\bar
\Pi}_n \left( {\bar{\Pi}}_n - \sum_{k=1}^N E_k^{(n)} \right)^{1/2}.
\end{equation} Clearly, $V_{\ul{j}}^{(n)} \leq {\bar{\Pi}}_n - \sum_{k=1}^N
E_k^{(n)}$, and we also have:

\begin{lemma} There exists an $n_1 \in \mathbb{N}$ such that 
if $n \ge n_1$ then  
\begin{equation}
\tr ({\bar \sigma}^{(n)} V_{\ul{j}}^{(n)}) \leq
2^{-n[{\chi}^*(\Phi) - \frac{2}{3}\epsilon]},\end{equation}
for all $\ul{j}$.
\label{lem47}
\end{lemma}
\textbf{Proof.} Let
$Q_n = \sum_{k=1}^{N(n)} E_k^{(n)}$.
Note that $Q_n$ is of the form $Q_n= {\tilde{Q}}_{ml_0} \otimes {\iden}_{n-ml_0}$,
since $E_j^{(n)} = {\cal{E}}_j^{ml_0} \otimes {\iden}_{n-ml_0}$. Note that
$Q_n$ commutes with ${\bar{\Pi}}_n$ by condition (i). Now, by
Lemma~\ref{lem44}, we have
\begin{equation} {\bar P}_{ml_0} {\bar \sigma}_{ml_0} {\bar P}_{ml_0}
\leq 2^{-m [S_M({\bar \phi}_\infty)- \frac{1}{4}\epsilon]} {\one}^{(ml_0)}
\end{equation} and, assuming that $l_0 \geq L$, we have by
Lemma~\ref{lem24}, 
\begin{eqnarray} {\bar P}_{ml_0} {\bar
\sigma}_{ml_0} {\bar P}_{ml_0} &\leq& 2^{-m[S(\Phi^{(l_0)}({\bar \rho}))
- \frac{1}{4}(1+\half l_0)\epsilon]} \one^{(ml_0)} \non \\ &\leq &
2^{-n[\frac{1}{l_0} S(\Phi^{(l_0)}({\bar \rho})) - \frac{1}{4}
\epsilon]} \one^{(ml_0)}
\end{eqnarray} provided \begin{equation} \frac{n-ml_0}{l_0}
S(\Phi^{(l_0)}({\bar \rho})) \leq \frac{1}{4} (n-m-\half ml_0) \epsilon,
\end{equation} which holds if $l_0 \geq 6$ and $$ m \geq
\frac{12}{\epsilon} \log\,\mbox{dim}\,{\cal K} $$ since $\frac{1}{l_0}
S(\Phi^{(l_0)}({\bar \rho})) \leq \log\,\mbox{dim}\,{\cal K}$.

Using this, we get
\begin{eqnarray} \lefteqn{\tr ( {\bar \sigma}^{(n)} V_{\ul{j}}^{(n)})} 
\non \\ &=&
\tr \left[ {\bar \sigma}^{(n)} ({\bar{\Pi}}_n - Q_n)^{1/2} {\bar
\Pi}_n {\Pi}_{\ul{j}}^{(n)} {\bar{\Pi}}_n ({\bar{\Pi}}_n - Q_n)^{1/2} \right]
\non
\\ &=& \tr \left[ {\bar{\Pi}}_n {\bar \sigma}^{(n)} {\bar{\Pi}}_n
({\bar{\Pi}}_n - Q_n)^{1/2} {\Pi}_{\ul{j}}^{(n)} ({\bar{\Pi}}_n - Q_n)^{1/2}
\right] \non \\ &=& \tr \left[ {\bar P}_{ml_0} {\bar \sigma}_{ml_0} {\bar P}_{ml_0}
({\bar P}_{ml_0} - {\tilde{Q}}_{ml_0})^{1/2} {P}_{\ul{j}}^{(ml_0)} ({\bar P}_{ml_0} - 
{\tilde{Q}}_{ml_0})^{1/2}
\right] \non \\
&\leq & 2^{-n[\frac{1}{l_0} S(\Phi^{(l_0)}({\bar
\rho}))-\frac{1}{4} \epsilon]} \tr \left[ ({\bar P}_{ml_0} - {\tilde{Q}}_{ml_0})^{1/2} {P}_{\ul{j}}^{(ml_0)} ({\bar P}_{ml_0} - 
{\tilde{Q}}_{ml_0})^{1/2} \right] \non \\ &\leq &
2^{-n[\frac{1}{l_0} S(\Phi^{(l_0)}({\bar \rho}))-\frac{1}{4}
\epsilon]} \tr\, ({P}_{\ul{j}}^{(ml_0)}).
\end{eqnarray}
However, by Lemma~\ref{lem46}, \begin{eqnarray}
\tr\,({P}_{\ul{j}}^{(ml_0)})&\leq & 
2^{m[{\bar S}_M(\{p_j^{(l_0)},\rho_j^{(l_0)}\}) + \frac{1}{4} \epsilon]}
\nonumber\\
&\leq & 2^{n[\frac{1}{l_0} {\bar
S}_M(\{p_j^{(l_0)},\rho_j^{(l_0)}\}) + \frac{1}{4} \epsilon]} \non \\
&\leq & 2^{n[\frac{1}{l_0} \sum_j p_j^{(l_0)} S(\rho_j^{(l_0)}) +
\frac{1}{4} \epsilon]},
\label{428}
\end{eqnarray} where the
last inequality follows from the subadditivity of the von Neumann 
entropy. The lemma now follows from \reff{chiapprox}. \qed

Since $N(n)$ is maximal it follows that
\begin{equation} \tr \left( \sigma_{\ul{j}}^{(n)} V_{\ul{j}}^{(n)}
\right) \leq 1-\epsilon. \label{Vj} \end{equation}

\begin{corollary} 
\begin{equation} {\mathbb{E}} \left( \tr \left[
\sigma_{\ul{j}}^{(n)} V_{\ul{j}}^{(n)} \right] \right) <
1-\epsilon.
\end{equation} \end{corollary}

\begin{lemma} Assume $\eta > 3 \delta$. Then for all $n \geq n_2= m_1 l_0 \vee m_2 l_0$, 
\begin{equation} {\mathbb{E}} \left( \tr
\left[ \sigma_{\ul{j}}^{(n)} {\bar{\Pi}}_n {\Pi}_{\ul{j}}^{(n)} {\bar{\Pi}}_n 
\right] \right) > 1-\eta. \end{equation} \label{lem48}
\end{lemma}
\textbf{Proof.} We write \begin{eqnarray} \lefteqn{ {\mathbb{E}} \left(
\tr \left[ \sigma_{\ul{j}}^{(n)} {\bar{\Pi}}_n {\Pi}_{\ul{j}}^{(n)}
{\bar{\Pi}}_n \right] \right) = } \non \\ &=&  {\mathbb{E}} \left( \tr
\left[ \sigma_{\ul{j}}^{(n)} {\Pi}_{\ul{j}}^{(n)} \right] \right) -
{\mathbb{E}} \left( \tr \left[ \sigma_{\ul{j}}^{(n)} (\one - {\bar{\Pi}}_n)
{\Pi}_{\ul{j}}^{(n)} \right] \right) \non \\ && \qquad - {\mathbb{E}} \left(
\tr \left[ \sigma_{\ul{j}}^{(n)} {\bar{\Pi}}_n {\Pi}_{\ul{j}}^{(n)}
(\one - {\bar{\Pi}}_n) \right] \right). \label{k0}\end{eqnarray} 

The first term equals ${\mathbb{E}}\bigl[\tr\bigl( \sigma_{\ul{j}}^{(ml_0)}{P}_{\ul{j}}^{(ml_0)}\bigr)\bigr]$,
which by Lemma~\ref{lem46} is $> 1-\delta^2$, provided $n
\geq n_2$. 

Note that
\begin{equation}
{\mathbb{E}} \left( \tr \left[ \sigma_{\ul{j}}^{(n)} (\one - {\bar{\Pi}}_n)
{\Pi}_{\ul{j}}^{(n)} \right] \right)=
{\mathbb{E}} \left( \tr \left[ \sigma_{\ul{j}}^{ml_0} (\one - {\bar P}_{ml_0})
{P}_{\ul{j}}^{(ml_0)} \right] \right)
\label{k1}
\ee
and similarly
\begin{equation}
{\mathbb{E}} \left(
\tr \left[ \sigma_{\ul{j}}^{(n)} {\bar{\Pi}}_n {\Pi}_{\ul{j}}^{(n)}
(\one - {\bar{\Pi}}_n) \right] \right)
={\mathbb{E}} \left(
\tr \left[ \sigma_{\ul{j}}^{(ml_0)} {\bar P}_{ml_0} {P}_{\ul{j}}^{(ml_0)}
(\one - {\bar P}_{ml_0}) \right] \right)
\label{k2}
\ee
Using \reff{k1} and \reff{k2}, the last two terms on the right hand side 
of \reff{k0} can be bounded using Cauchy-Schwarz
and Lemma~\ref{lem44} as follows :
\begin{equation} {\mathbb{E}} \left( \tr \left[
\sigma_{\ul{j}}^{(n)} (\one - {\bar{\Pi}}_n) {\Pi}_{\ul{j}}^{(n)} \right]
\right) \leq \delta
\end{equation} and
\begin{equation} {\mathbb{E}} \left( \tr \left[
\sigma_{\ul{j}}^{(n)} {\bar{\Pi}}_n {\Pi}_{\ul{j}}^{(n)} (\one - {\bar{\Pi}}_n) \right] \right) \leq \delta
\end{equation} provided $n \geq n_1$.
Choosing $n_3 = n_1 \vee n_2$ and $\delta^2 + 2 \delta < \eta$ the
result follows. 
\qed

\begin{lemma} Assume $\eta < \frac{1}{3} \epsilon$ and $\eta > 3\delta$. Then for $n \geq n_3= n_1 \vee n_2$,
\begin{equation} \tr
\left[ {\bar\sigma}_n \sum_{k=1}^N E_k \right] = {\mathbb{E}} \left( \tr
\left[ \sigma_{\ul{j}}^{(n)} \sum_{k=1}^N E_k \right] \right) \geq
\eta^2. \end{equation}  \label{lemma9}
\end{lemma}
\textbf{Proof.} Define \begin{equation} Q'_n = {\bar{\Pi}}_n -
({\bar{\Pi}}_n-Q_n)^{1/2}. \end{equation} By the above corollary,
\begin{eqnarray} 1-\epsilon &\geq& {\mathbb{E}} \left\{ \tr \left(
\sigma_{\ul{j}}^{(n)} ({\bar{\Pi}}_n-Q'_n) {\Pi}_{\ul{j}}^{(n)} ({\bar
P}_n-Q'_n) \right) \right\} \non \\ &=& {\mathbb{E}} \left\{ \tr \left(
\sigma_{\ul{j}}^{(n)} {\bar{\Pi}}_n {\Pi}_{\ul{j}}^{(n)} {\bar{\Pi}}_n
\right) \right\} \non \\ && - {\mathbb{E}} \left\{ \tr \left(
\sigma_{\ul{j}}^{(n)} Q'_n {\Pi}_{\ul{j}}^{(n)} {\bar{\Pi}}_n \right)  +
\tr \left( \sigma_{\ul{j}}^{(n)} {\bar{\Pi}}_n {\Pi}_{\ul{j}}^{(n)}
Q'_n \right) \right\} \non \\ && + {\mathbb{E}} \left\{ \tr \left(
\sigma_{\ul{j}}^{(n)} Q'_n {\Pi}_{\ul{j}}^{(n)} Q'_n \right) \right\}.
\end{eqnarray} Since the last term is positive, we have, by
Lemma~\ref{lem48},
\begin{equation} {\mathbb{E}} \left\{ \tr \left( \sigma_{\ul{j}}^{(n)}
Q'_n {\Pi}_{\ul{j}}^{(n)} P_n \right)  + \tr \left(
\sigma_{\ul{j}}^{(n)} P_n {\Pi}_{\ul{j}}^{(n)} Q'_n \right) \right\}
\geq \epsilon - \eta > 2 \eta.
\end{equation}  On the other hand,
using Cauchy-Schwarz for each term, the left-hand side is bounded
by
\begin{equation} 2\left\{ {\mathbb{E}} \left[ \tr \left(
\sigma_{\ul{j}}^{(n)} Q^{\prime 2}_n \right) \right]
\right\}^{1/2}. \end{equation} Thus,
\begin{equation}  {\mathbb{E}} \left[ \tr \left( \sigma_{\ul{j}}^{(n)}
Q^{\prime 2}_n \right) \right] \geq \eta^2.
\end{equation} To complete the proof, we now claim that
\begin{equation} Q_n \geq (Q'_n)^2. \end{equation} Indeed, this
follows on the domain of $P_n$ from the inequality $1-(1-x)^2 \geq
x^2$ for $0 \leq x \leq 1$. \qed

To complete the proof of the theorem, we now have by assumption,
\begin{equation} \tr \left[ {\bar \sigma}^{(n)} E_k^{(n)} \right] \leq
2^{-n[\chi^*(\Phi) - \frac{2}{3} \epsilon]}
\end{equation} for all $k=1,\dots,N(n)$. On the other hand, choosing
$\eta < \frac{1}{3} \epsilon$ and $\delta < \third \eta$, we have by
Lemma~\ref{lemma9},
\begin{equation} \tr
\left[ {\bar\sigma}^{(n)} \sum_{k=1}^N E_k \right] \geq \eta^2
\end{equation} provided $n \geq n_3$.
It follows that \begin{equation} N(n) \geq \eta^2
2^{n [\chi^*(\Phi) - \frac{2}{3} \epsilon]} \geq 2^{n
[\chi^*(\Phi) - \epsilon]} \end{equation} for $n \geq
n_3$ and $n \geq -\frac{6}{\epsilon} \log \eta.$ \qed

\section{The case of a general Markov chain}
\label{general}

In the following  we write 
\begin{equation} \Phi_{C,i}^{(n)} = \Phi_{i_k} \otimes \dots \otimes \Phi_{i_{k+n-1}}
\end{equation} if $i=i_k$ is in a periodic class $C$, and where the labelling is modulo the 
length of the class. 

\subsection{The direct part of Theorem \ref{main1}}
In this section we prove the direct part of Theorem \ref{main1}.
As in the ergodic case (considered in Section \ref{ergodic}),
we once again employ a quantum Feinstein Lemma, which is a
generalization of Lemma \ref{feinmem} and is given by the following lemma.

\begin{lemma} \label{fein_markov}
For all $\epsilon > 0$, there exists $n_0 \in {\mathbb{N}}$
such that there exist at least $N =N_n \geq 2^{n[C(\Phi) -
\epsilon]}$ product states ${\tilde \rho}_1^{(n)}, \dots, {\tilde
\rho}_N^{(n)}$ on ${\cal H}^{\otimes n}$ and positive operators
$E_1^{(n)}, \dots, E_N^{(n)}$ on ${\cal K}^{\otimes n}$ such that
$\sum_{k=1}^N E_k^{(n)} \leq \one$ and \begin{equation} \Tr \left[
\Phi^{(n)} ({\tilde \rho}_k^{(n)}) \, E_k^{(n)} \right] > 1-\epsilon
\label{inderr} \end{equation} for all $k=1,\dots,N$. \end{lemma}

The proof of this lemma is given in Section \ref{proof}. It uses
the idea of adding a preamble to the codewords (as was done in \cite{DD1})
to distinguish between the different classes of the Markov chain.
The construction of the preamble is discussed in detail in the following section. 

\subsection{Construction of a preamble}

To distinguish between the different classes,
$\Phi_C^{(n)}$, of the quantum channel $\Phi$, we add a preamble to the
input state encoding each message in the set ${\cal{M}}_n$. This is
given by an $m$-fold tensor product of suitable states (as
described below). Let us first sketch the idea behind adding such
a preamble. Helstr\o m \cite{helstrom} showed that two states
$\sigma_1$ and $\sigma_2$, occurring with {\em{a priori}} probabilities
$\gamma_1$ and $\gamma_2$ respectively, can be distinguished with
an asymptotically vanishing probability of error, if a suitable
collective measurement is performed on the $m$-fold tensor
products $\sigma_1^{\otimes m}$ and $\sigma_2^{\otimes m}$, for a
large enough $m \in {\mathbb{N}}$. The optimal measurement is
projection-valued. The relevant projection operators, which we
denote by $\Pi^+$ and $\Pi^-$, are the orthogonal projections onto
the positive and negative eigenspaces of the difference operator
$A_m= \gamma_1\sigma_1^{\otimes m} - \gamma_2\sigma_2^{\otimes
m}$. Here we generalize this result to distinguish between the
different classes $\Phi_C^{(n)}$. If the preamble is given by a state
$\omega^{\otimes m}$, then, by using Helstr\o m's result, we can
construct a POVM which distinguishes between the output states
$\sigma_C^{(n)}:= \Phi_C^{(n)}(\omega^{\otimes m})$
corresponding to the different classes $\Phi_C^{(n)}$.
The outcome of this POVM measurement would in turn serve to
determine which class of the channel is being used for
transmission.

We first show that there exists a preamble that can
distinguish between the different classes, analogous to the
branches in \cite{DD1}. In fact, we want to do more. In the case of
periodic classes, we also want to distinguish between initial
states of the class. We therefore subdivide the problem into the following
four possibilities: \begin{enumerate} \item To distinguish between
two aperiodic classes; \item To distinguish between an aperiodic
class $C$ and an initial state $i'$ of a periodic class $C'$;
\item To distinguish between two periodic classes $C$ and $C'$;
and \item To distinguish between the states of a single periodic
class. \end{enumerate} 

We refer to the aperiodic classes and the periodic classes with
given initial state, as {\em{branches}} of the channel.

Consider the first problem: distinguishing
between two aperiodic classes. We can obviously assume that the
$\Phi_C^{(n)} \neq \Phi^{(n)}_{C'}$ for some $n$: otherwise the
classes are identical and we can combine their probabilities. This
means that for any pair of aperiodic classes $C,C'$ there exists
$n = n(C,C')$ and a state $\omega_{C,C'}^{(n)}$ such that
$\Phi_C^{(n)}(\omega_{C,C'}^{(n)}) \neq
\Phi_{C'}^{(n)}(\omega_{C,C'}^{(n)})$. 
In fact, in most cases we
can take $n=1$, and we shall assume this for simplicity in the
following, even though this is not necessary.

Introducing the fidelity of two states as in \cite{Nielsen},
\begin{equation} F(\sigma,\sigma') = \Trace \sqrt{\sigma^{1/2}
\sigma'\, \sigma^{1/2}}, \end{equation} we then have
\begin{equation} 
F(\Phi_C^{(1)}(\omega_{C,C'}),
\Phi_{C'}^{(1)}(\omega_{C,C'})) \leq f < 1,
\ee
for all pairs $C,C'$ with $C < C'$ in some arbitrary ordering
of ${\cal C}_{\rm aper}$, the set of aperiodic classes.

The following lemma shows that the classes $C$ and $C'$ can be distinguished.
\begin{lemma} For any two aperiodic classes $C$ and $C'$,
\begin{equation} F( \Phi^{(m)}_C (\omega_{C,C'}^{\otimes m}),
\Phi^{(m)}_{C'}(\omega_{C,C'}^{\otimes m})) \to 0 \mbox{ as } m
\to \infty. \end{equation} \label{L3.1} \end{lemma}
\textbf{Proof.} Choose $\alpha > 0$ so small that $1+\alpha <
f^{-1}$. First let $k$ be so large that
\begin{equation} (1-\alpha) \gamma_j < \sum_{i_2,\dots,i_{k-1}} q_{ii_2} \dots
q_{i_{k-1}j} <(1+ \alpha) \gamma_{j} \label{equilib}
\end{equation} for all $i,j \in I$. Now let $\{{\cal{E}}_r\}_r$ 
be a POVM such that 
\begin{equation} F(\sigma, \sigma') = \sum_r \sqrt{\Tr(\sigma
{\cal{E}}_r)\,\Tr(\sigma' {\cal{E}}_r)}, \end{equation} 
(see e.g. Eq.(9.74) in \cite{Nielsen}) where we denote
\begin{equation} \sigma = \Phi_C^{(1)}(\omega_{C,C'}) \mbox{ and }
\sigma' = \Phi_{C'}^{(1)}(\omega_{C,C'}). \end{equation} Then we
have \begin{eqnarray} \lefteqn{F \left( \Phi^{(mk+m)}_C
(\omega_{C,C'}^{\otimes (mk+m)}),
\Phi^{(mk+m)}_{C'}(\omega_{C,C'}^{\otimes (mk+m)}) \right) \leq }
\non
\\ &\leq & \sum_{r_1,\dots,r_m} \left[ \Tr \left(
\Phi^{(mk+m)}_{C}(\omega_{C,C'}^{\otimes (mk+m)})
\bigotimes_{i=1}^m ({\cal{E}}_{r_i} \otimes \one_k) \right) \right. \non
\\ && \qquad \times \left. \Tr \left(
\Phi^{(mk+m)}_{C'}(\omega_{C,C'}^{\otimes (mk+m)})
\bigotimes_{i=1}^m ({\cal{E}}_{r_i} \otimes \one_k) \right) \right]^{1/2}
\non \\ &\leq & \sum_{r_1,\dots,r_m} (1+\alpha)^{m-1}
\prod_{i=1}^m \sqrt{\Tr(\sigma {\cal{E}}_{r_i})\,\Tr(\sigma' {\cal{E}}_{r_i})}
\non \\ &=& (1+\alpha)^{m-1} F(\sigma, \sigma')^m \to 0.
\end{eqnarray} \hfill \qed

Next consider the second case, i.e., to distinguish an aperiodic
class $C$ and an initial state $i'$ of a periodic class $C'$.
There exists a state $\omega = \omega_{C,i'}$ on $\cal H$ such
that \begin{equation} f := F(\Phi^{(1)}_C(\omega_{C,i'}),
\Phi_{i'}(\omega_{C,i'})) < 1. \end{equation}

\begin{lemma} Let $C$ be an aperiodic class and $C'$ a periodic
class with length $L=L(C')$, let $i' \in C'$, and choose $\omega =
\omega_{C,i'}$ as above. Then \begin{equation} F \left(
\Phi_C^{(m)}(\omega^{\otimes m} ) ,
\Phi_{C',i'}^{(m)}(\omega^{\otimes m}) \right) \to 0
\end{equation} as $m \to \infty$. \label{L3.2} \end{lemma}
\textbf{Proof.} We proceed as in Lemma~\ref{L3.1} and choose
$\alpha > 0$ so small that $1+\alpha < f^{-1}$ and let $k$ be so
large that (\ref{equilib}) holds and in addition such that $k$
is a multiple of $L$. Again, we let $\{{\cal{E}}_r\}_r$ be a POVM such
that \begin{equation} F(\sigma, \sigma') = \sum_r \sqrt{\Tr(\sigma
{\cal{E}}_r)\,\Tr(\sigma' {\cal{E}}_r)}, \end{equation} where now \begin{equation}
\sigma = \Phi_C^{(1)}(\omega_{C,i'}) \mbox{ and } \sigma' =
\Phi_{i'}( \omega_{C,i'}). \end{equation} Then \begin{eqnarray}
\lefteqn{F \left( \Phi^{(mk+m)}_C (\omega_{C,i'}^{\otimes
(mk+m)}), \Phi^{(mk+m)}_{i'}(\omega_{C,i'}^{\otimes (mk+m)})
\right) } \non
\\ &\leq & \sum_{r_1,\dots,r_m} \left[ \Tr \left(
\Phi^{(mk+m)}_{C}(\omega_{C,C'}^{\otimes (mk+m)})
\bigotimes_{i=1}^m ({\cal{E}}_{r_i} \otimes \one_k) \right) \right. \non
\\ && \qquad \times \left. \prod_{j=1}^m \Tr \left(
\Phi_{i'}(\omega_{C,i'}) \,{\cal{E}}_{r_j} \right) \right]^{1/2} \non \\
&\leq & \sum_{r_1,\dots,r_m} (1+\alpha)^{\frac{m-1}{2}}
\prod_{j=1}^m \sqrt{\Tr(\sigma {\cal{E}}_{r_j})\,\Tr(\sigma' {\cal{E}}_{r_j})}
\non
\\ &=& (1+\alpha)^{\frac{m-1}{2}} F(\sigma, \sigma')^m \to 0 \quad {\hbox{as}} \,\, 
m \rightarrow \infty.
\end{eqnarray} \hfill \qed

Distinguishing two periodic classes is straightforward:

\begin{lemma} If $C$ and $C'$ are two different periodic classes
with periods $L(C)$ and $L(C')$ respectively, then there exists a
state $\omega_{C,C'}^{(L)}$ on ${\cal H}^{\otimes L}$, where $L =
L(C) \, L(C')$ such that \begin{equation} F \left(
\Phi_C^{(mL)}((\omega_{C,C'}^{(L)})^{\otimes m}) ,
\Phi_{C'}^{(mL)}((\omega_{C,C'}^{(L)})^{\otimes m}) \right) \to 0
\end{equation} as $m \to \infty$. \label{L3.3} \end{lemma}
\textbf{Proof.} Since the two periodic classes are distinct, there
exists a state $\omega = \omega_{C,C'}^{(L)}$ such that
\begin{equation} \sigma = \Phi_C^{(L)}(\omega_{C,C'}^{(L)}) \neq
\Phi_{C'}^{(L)}(\omega_{C,C'}^{(L)}) \end{equation} (In fact we
can take $L$ to be the least common multiple of $L(C)$ and
$L(C')$.) Then writing $\omega = \omega_{C,C'} \otimes \varphi^{\otimes k} $, 
where $\varphi$ is an arbitrary state on $\cal H$ and $k$ is so large that 
(\ref{equilib}) holds, \begin{eqnarray} \lefteqn{F \left(
\Phi_C^{(mL+mk)}(\omega^{\otimes m}),
\Phi_{C'}^{(mL)}(\omega^{\otimes m}) \right) \leq } \non \\ &\leq & 
(1+\alpha)^m F \left( \left(
\Phi_C^{(L)}(\omega) \right)^{\otimes m}, \left(
\Phi_{C'}^{(L)}(\omega) \right)^{\otimes m} \right) \to 0.
\end{eqnarray} \qed

Finally, to distinguish the initial states of a given periodic
class $C$, notice first of all that the corresponding CPT maps
$\Phi_i$ \emph{need not all be distinct!} However, we may assume that
there is no internal periodicity of these maps within a periodic
class; otherwise the class can be contracted to a single such
period. This means, that for any two states $i,i' \in C$ there
exists $l \leq L(C)-1$ such that $\Phi_{i+l} \neq \Phi_{i'+l}$.
Then choose $\omega = \omega_{i,i'}$ such that \begin{equation} f
:= F(\Phi_{i+l}(\omega), \Phi_{i'+l}(\omega)) < 1. \end{equation}

\begin{lemma} If $C$ is a periodic class with period $L(C)$, $i,i'
\in C$ and $\omega$ is a state as above, then \begin{equation} F
\left( \Phi^{(m)}_{C,i}(\omega^{\otimes m}),
\Phi_{C,i'}^{(m)}(\omega^{\otimes m}) \right) \to 0 
\quad {\hbox{as}}\,\, m \to \infty \end{equation}
\label{L3.4} \end{lemma}
\textbf{Proof.} \begin{eqnarray} \lefteqn{ F \left(
\Phi^{(m)}_{C,i}(\omega^{\otimes m}), \Phi_{C,i'}^{(m)}(\omega^{\otimes
m}) \right) } \non \\ &=& \left[ F \left(
\Phi_{C,i}^{(L)}(\omega^{\otimes L}), \Phi_{C,i'}^{(L)} (\omega^{\otimes
m}) \right) \right]^m \non \\ &\leq & \left[ F(\Phi_{i+l}(\omega),
\Phi_{i'+l}(\omega)) \right]^m = f^m \to 0. \end{eqnarray} \qed

%-----------------------------------------------------------------

We now introduce, in each of the four cases, difference operators 
$A_{C,C'}^{(m)}$, $A_{C,i'}^{(m)}$, $A_{i,i'}^{(m)}$ with $i,i'$ in a 
periodic class, and corresponding projections $\Pi^\pm_{C,C'}$, $\Pi^\pm_{C,i'}$ 
and $\Pi^\pm_{i,i'}$ onto their positive and negative eigenspaces , which serve 
to distinguish the different possibilities, as in \cite{DD1}. The difference 
operators are defined by  
\begin{equation} A^{(m)}_{C,C'} = \gamma_C
\bigl(\Phi_C^{(m)}(\omega_{C,C'})\bigr)^{\otimes m} - \gamma_{C'}
\bigl(\Phi_{C'}^{(m)}(\omega_{C,C'})\bigr)^{\otimes m}, \end{equation} 
\begin{equation} A^{(m)}_{C,i'} = \gamma_C
\bigl(\Phi_C^{(m)}(\omega_{C,i'})\bigr)^{\otimes m} - \gamma_{i'}
\bigl(\Phi_{C'}^{(m)}(\omega_{C,i'})\bigr)^{\otimes m}, \end{equation}
and 
\begin{equation} A^{(m)}_{i,i'} = \gamma_i
\bigl(\Phi_{C,i'}^{(m)}(\omega_{i,i'})\bigr)^{\otimes m} - \gamma_{i'}
\bigl(\Phi_{C,i'}^{(m)}(\omega_{i,i'})\bigr)^{\otimes m}. \end{equation}
The following lemma was proved in \cite{DD1}:

\begin{lemma} \label{LPi} Suppose that for a given $\delta >0$,
\begin{equation} |\tr
[|A_{c,c'}^{(m)}|] - (\gamma_c + \gamma_{c'})| \leq \delta.
\end{equation} Then \begin{equation}
|\tr [\Pi_{c,c'}^+ \bigl(\Phi_c^{(m)}(\omega_{c,c'}^{\otimes m})\bigr)]
- 1| \leq \frac{\delta}{2 \gamma_c}
\end{equation} and
\begin{equation}
|\tr [\Pi_{c,c'}^- \bigl(\Phi_{c'}^{(m)}(\omega_{c,c'}^{\otimes m})\bigr)]-1| \leq \frac{\delta}{2 \gamma_{c'}}. \end{equation}
\end{lemma}

Here $c,c'$ denote either two different classes $C,C'$ or one aperiodic class 
$C$ and an initial state $i'$ in a periodic class, or two different initial 
states in the same periodic class.

To compare the outputs of all the different branches of the
channel, we define projections ${\tilde \Pi}_i$ on the tensor
product space ${\cal K}^{\otimes mM}$ where 
\be
M= M_1 + M_2 + M_3 + M_4,
\label{em}
\ee 
with 
\begin{enumerate}
\item $M_1$ is the total number 
of pairs of aperiodic classes;
\item $M_2$ is the total number of pairs of periodic classes;
\item $M_3$ is the total number of pairs of aperiodic classes and intial states of 
periodic classes and
\item $M_4$ is the total number of pairs of states in the same periodic class.
\end{enumerate}
We introduce an arbitrary order on the classes $ C \in {\cal C}$ assuming
$C < C'$ if $C \in {\cal C}_{\rm aper}$ and $C' \in {\cal C}_{\rm per}$. Then we put
\begin{equation} {\tilde \Pi}_c = \bigotimes_{(c',c''):\, c'<c''} \Gamma_{c',c''}^{(c)}, 
\mbox{ where }
\Gamma_{c',c''}^{(c)} = \left\{ \begin{array}{lcl} I_m &\mbox{ if
} &c' \neq c \mbox{ and } c'' \neq c \\ \Pi_{c',c}^- &\mbox{ if } &c'' = c \\
\Pi_{c,c''}^+ &\mbox{ if } &c' = c. \end{array} \right.
\end{equation}

It follows from the fact that $\Pi_{c',c''}^+ \Pi_{c',c''}^-
= 0$, that the projections ${\tilde \Pi}_c$ are also disjoint:
\begin{equation} {\tilde \Pi}_{c_1} {\tilde \Pi}_{c_2} = 0 \quad
{\hbox{for }} \, c_1\ne c_2.
\end{equation}

We use the following lemma.
\begin{lemma} \label{Ldistinct}
For all aperiodic classes $C$, \begin{equation} \lim_{m \to \infty} \tr
\left[ {\tilde \Pi}_C\, \Phi_C^{(mM)} \left( \omega^{(mM)}
\right) \right] = 1, \end{equation} and for all periodic classes $C$ 
and all $i \in C$, 
\begin{equation} \lim_{m \to \infty} \tr
\left[ {\tilde \Pi}_{C,i}\, \Phi_{C,i}^{(mM)} \left( \omega^{(mM)}
\right) \right] = 1. \label{530}\end{equation} \end{lemma}
\textbf{Proof.} Notice that for all $(c,c')$, \begin{eqnarray}
&& F(\gamma_c \Phi_c^{(mM)}(\omega_{c,c'})^{\otimes m}, \gamma_{c'}
\Phi_{c'}^{(mM)}(\omega_{c,c'})^{\otimes m}) \non \\ && \qquad  
= \sqrt{\gamma_c \gamma_{c'}} F(\Phi_c^{(mM)}(\omega_{c,c'}), 
\Phi_{c'}^{(mM)}(\omega_{c,c'})) \to 0
\end{eqnarray} as $m \to \infty$. Using the inequalities \cite{Nielsen}
\begin{eqnarray} \tr (A_1) +
\tr (A_2) - 2 F(A_1,A_2) &\leq& || A_1 - A_2 ||_1 \non \\ && \qquad \leq \tr
(A_1) + \tr (A_2) \non \\ && \end{eqnarray} for any two positive operators 
$A_1$ and $A_2$, we find that
\begin{equation} \big|\,\tr\,\bigl(|A_{c,c'}^{(m)}|\bigr) -
(\gamma_i + \gamma_{j}) \big| \leq \delta_m,
\end{equation}
where $\delta_m \to 0$ as $m \to \infty$,
since
\begin{equation} \tr\,\bigl(|A_{c,c'}^{(m)}|\bigr) = ||
\gamma_c \Phi_{c}^{(mM)}(\omega_{c,c'})^{\otimes m} - \gamma_{c'}
\Phi_{c'}^{(mM)}(\omega_{c,c'})^{\otimes m} ||_1. \end{equation}
We now replace $m$ by $m'=m+k$, where $k \in \mathbb{N}$
is large enough so that \reff{equilib} holds, and 
define
\begin{equation} \omega^{(m'M)} :=
\bigotimes_{(c_1,c_2)} \omega_{c_1,c_2}^{\otimes (m+k)},
\label{preamble}
\end{equation}
Using \reff{equilib} to separate the different classes, 
we then have for any $C \in {\cal C}_{\rm aper}$,
\begin{eqnarray} 1 &\geq & \tr \left[ {\tilde \Pi}_C
\Phi_C^{(m'M)} \left( \bigotimes_{c_1 < c_2}
\omega_{c_1,c_2}^{\otimes (m+k)} \right) \right]  \non \\ &\geq &
(1-\alpha)^M \prod_{C' \in {\cal C}_{\rm aper}; C' < C} \tr \left[ \Pi_{C',C}^-
\bigl(\Phi_C^{(m)}(\omega_{C',C}^{\otimes m})\bigr) \right] \non \\ && \qquad \qquad \times
\prod_{C'' \in {\cal C}_{\rm aper}; C'' > C} \tr \left[ \Pi_{C,C''}^+
\bigl(\Phi_C^{(m)}(\omega_{C,C''}^{\otimes m})\bigr) \right] \non \\ && \qquad \qquad \times
\prod_{C' \in {\cal C}_{\rm per}} \prod_{i' \in C'} \tr \left[ \Pi_{C,i'}^+
\bigl(\Phi_C^{(m)}(\omega_{C,i'}^{\otimes m})\bigr) \right] \non \\
&\geq& (1-\alpha)^M \left(1-\frac{\delta_m}{2\gamma_C}\right)^{|{\cal C}_{\rm aper}|-1 +
\sum_{C' \in {\cal C}_{\rm per}} |C'|} \to 1, \end{eqnarray}
since $\delta_m \to 0$ as $m \to \infty$. The last inequality follows from Lemma~\ref{LPi}.
\\ 

The analogous result, \reff{530}, for periodic classes, is proved in
a similar manner. \qed

\vskip1cm

\subsection{Proof of Lemma~\ref{fein_markov}}
\label{proof}

Given $\delta > 0$, we now fix $m_0$ so large that 
\begin{equation}  \tr \left[
{\tilde \Pi}_C \,\Phi_C^{(m_0 M)} \left( \omega^{(m_0 M)} \right)
\right] > 1-\delta \label{Piaper} \end{equation} for all
$C \in {\cal C}_{\rm aper}$ and \begin{equation} \tr \left[
{\tilde \Pi}_{C',i'} \,\Phi_{C',i'}^{(m_0 M)} \left( \omega^{(m_0 M)} \right)
\right] > 1-\delta \label{Piper} \end{equation} for all 
$C' \in {\cal C}_{\rm per}$ and $i' \in C'$.
Here $M$ is given by \reff{em}. 
The product state $ \omega^{(m_0 M)}$, defined through \reff{preamble}, 
is used as a preamble to the input state encoding
each message, and serves to distinguish between the different
branches of the channel, i.e., between $\Phi_C$, $C \in {\cal C}_{aper}$ and 
$\Phi_{C',i}$, $C' \in {\cal C}_{per}$ and $i \in C'$.
If $\rho_k^{(n)} \in {\cal{B}}({\cal{H}}^{\otimes n})$ is a 
state encoding the $k^{th}$ classical message in the set
${\cal{M}}_n$, then the $k^{th}$ codeword is given by the product
state $$ \omega^{(m_0 M)} \otimes \rho_k^{(n)} .$$
 
We follow the same steps as in the proof of Theorem
5.1 in \cite{DD1}. First we fix $l_0$ large enough, and an ensemble
$\{p_j^{(l_0)},\rho_j^{(l_0)}\}$ such that 
\begin{equation} \left| C(\Phi) -
\bigwedge_{C \in {\cal C}} {\bar{\chi}}^{(l_0)}_C (\{p_j^{(l_0)},\rho_j^{(l_0)}\}) \right| <
\frac{\epsilon}{6}. \label{540}\end{equation}

%-----------------------------------------------------------------------------

As in the ergodic case (Section \ref{ergodic}), let
$N={\tilde N}(n)$ be the maximal number of product states ${\tilde
\rho}_1^{(n)}, \dots, {\tilde \rho}_N^{(n)}$ on ${\cal H}^{\otimes
n}$  for which there exist positive
operators $E_1^{(n)}, \dots, E_N^{(n)}$ on ${\cal K}^{\otimes m_0 M}
\otimes {\cal K}^{\otimes n}$ such that

\begin{enumerate}
\item[(i)] $  E_k^{(n)} = \sum_{C \in {\cal C}_{\rm aper}} {\tilde \Pi}_C \otimes
E_{k,C}^{(n)} + \sum_{C' \in {\cal C}_{\rm per}} \sum_{i' \in C'} {\tilde \Pi}_{C',i'} 
\otimes E_{k,i'}^{(n)}$ and \newline $ \sum_{k=1}^N E_{k,C}^{(n)} \leq {\bar
P}_{C,n}$;  $ \sum_{k=1}^N E_{k,i'}^{(n)} \leq {\bar
P}_{i',n}$ for $i' \in C' \in {\cal C}_{\rm per}$, and 
\item[(ii)] $ \disp{ \sum_{C \in {\cal C}_{\rm aper}} \gamma_C \tr
\left[\,({\tilde \Pi}_C \otimes E_{k,C}^{(n)}) \Phi_C^{(m_0 M + n)}
\left( \omega^{(m_0 M)} \otimes {\tilde\rho}_k^{(n)} \right)  \right]} $ \newline 
$ \disp{ + \sum_{C' \in {\cal C}_{\rm per}} \sum_{i' \in C'} \gamma_{i'} \tr
\left[\,({\tilde \Pi}_{C',i'} \otimes E_{k,C}^{(n)}) \Phi_{C',i'}^{(m_0 M + n)}\left( \omega^{(m_0 M)}  \otimes {\tilde\rho}_k^{(n)} \right) \right]} $ \newline \hfill $> 1-\epsilon$, and 
\item[(iii)] $ \disp{ \sum_{C \in {\cal C}_{\rm aper}} \gamma_C \tr \left[\,({\tilde
\Pi}_C \otimes E_{k,C}^{(n)} ) \Phi_C^{(m_0 M + n)} \left( \omega^{(m_0 M)} \otimes 
{\bar \rho}^{(n)} \right) \right] } $ \newline $ + 
\disp{ \sum_{C' \in {\cal C}_{\rm per}} \sum_{i' \in C'} \gamma_{i'} 
\tr \left[\, ({\tilde \Pi}_{C',i'} \otimes E_{k,i'}^{(n)}) \Phi_C^{ (m_0 M + n)}
\left( \omega^{(m_0 M)} \otimes {\bar \rho}^{(n)} \right) \right]}$ \newline 
\hfill $ \leq 2^{-n[C(\Phi) - \half \epsilon]} $.
\end{enumerate}

Note that, as in the ergodic case, we can append ${\one}^{(n-ml_0)}$ to all POVM elements,
to reduce the proof to the case $n=ml_0$. In the following we therefore assume
$n=ml_0$ for simplicity.

The typical projection ${\bar P}_{C,n}$ for an aperiodic class is defined 
as before by Lemma \ref{lem44}. For a periodic class $C'$ we define the typical spaces
by interlacing those for the product channels $\Phi_i^{\otimes n}$ ($i \in C'$), as follows:

\begin{lemma} \label{pertyp} Let $C'$ be a periodic class with period $L$. 
Given $\epsilon, \delta > 0$, there exists $m_1' \in {\mathbb{N}}$ 
such that for $m \geq m'_1$ there are subspaces $\ol{\cal N}^{(n)}_{i,\epsilon} \subset {\cal K}^{\otimes m}_{l_0}$ ($i \in C'$),
$(n=ml_0)$, with projections ${\bar P}_{i,n}$ such that 
\begin{equation} {\bar P}_{i,n}  \Phi_{C',i} (\rho_{l_0}^{\otimes m}) {\bar P}_{i,n} \leq
2^{-m[S_{C'}-\frac{\epsilon}{4}]}, \end{equation}
where
$$S_{C'} = \frac{1}{L} \sum_{i=0}^{L-1}S(\Phi_{C',i}^{(l_0)}({\bar \rho}_{l_0})),$$ 
 
and \begin{equation} \tr \left( \Phi_{C',i}^{(n)}({\bar \rho}_{l_0}^{\otimes m}) 
{\bar P}_{i,n} \right) > 1-\delta^2. \end{equation} \end{lemma} 
\textbf{Proof.} We simply let  $\ol{\cal N}^{(n)}_{i,\epsilon}$ be the subspace 
spanned by the vectors $|\psi_{i,k_1}\rangle \otimes \dots \otimes |\psi_{i+l_0(n-1),k_n}\rangle$,
where $|\psi_{i,k}\rangle$ is an eigenvector of $\Phi_{C',i}^{(l_0)}$ and $|\psi_{i,k_1}\rangle \otimes |\psi_{i,k_{L+1}}\rangle \otimes \dots \otimes |\psi_{i,k_{[(n-1)/L]L+1}} \rangle$ belongs to the typical space for $\Phi_{C,i}^{(l_0)}({\bar \rho}_{l_0})$, $|\psi_{i+1,k_2}\rangle \otimes |\psi_{i+1,k_{L+1}}\rangle \otimes \dots \otimes |\psi_{i+1,k_{[(n-1)/L]L+2}} \rangle$  to that of $\Phi_{C,i+1}^{(l_0)}({\bar \rho}_{l_0})$, etc. \qed

Similarly we have:

\begin{lemma} \label{lem58} Let $C'$ be a periodic class with period $L$. 
Given $i \in C'$, and a sequence $\ul{j} = (j_1,\dots,j_m)\in \{1,2,\ldots ,J\}^m$, let $P_{i,\ul{j}}^{(n)}=P_{(C',i),\ul{j}}^{(n)}$ be the projection onto the subspace of ${\cal K}^{\otimes n}$ spanned by the eigenvectors of
$$\Phi_{C',i}^{(n)}(\rho^{(l_0)}_{\ul{j}}) = \bigotimes_{r=1}^m \Phi_{C',i+(r-1)l_0}^{(l_0)}(\rho_{j_r}^{(l_0)}),$$ 
with eigenvalues $\lambda_{\ul{j},\ul{k}} = 
\prod_{r=1}^m \lambda_{i+(r-1)l_0,j_r,k_r}$ such that 
\begin{equation} \left| \frac{1}{m} \log \lambda_{\ul{j},\ul{k}} + {\bar S}_{C'} \right| 
< \frac{\epsilon}{4}, \end{equation} where 
\begin{equation} {\bar S}_{C'} = \lim_{m \to \infty} 
\frac{1}{mL} \sum_{i' \in C'} \sum_{\ul{j}} p_{\ul{j}}^{(n)} S \left( 
(\Phi_{i'} \otimes \dots \otimes \Phi_{i'+ ml_0 -1}) (\rho_{j_1}^{(l_0)} \otimes \dots \otimes
\rho_{j_m}^{(l_0)}) \right). \end{equation} 
For any $\delta > 0$ there exists $m'_2$ such that for $m\geq m'_2$, 
\begin{equation} \mathbb{E} \left( \tr \left[ \Phi_{C',i}^{(ml_0)} \left(\bigotimes_{r=1}^m 
\rho_{j_r}^{(l_0)} \right) P_{i,\ul{j}}^{(n)} \right] \right) > 1-\delta^2. 
\end{equation} 
\end{lemma} 
Note that ${\bar S}_{C'}$ can be equivalently expressed as
$$
{\bar S}_{C'} = \lim_{m \to \infty}\frac{1}{mL}\sum_{i' \in C'} {\overline{S}}_{i'},
$$
with ${\overline{S}}_{i'}$ as in \reff{313}.

The remainder of the proof is identical to that of Theorem 5.1 in \cite{DD1}.
For each $c=C$ or $c=(C',i')$ with $i' \in C'\in  {\cal{C}}_{per}$, and $\uj = (j_1,\dots,j_m),$ we
define, as before \begin{equation} V_{c,\uj}^{(n)} = \left( {\bar
P}_c^{(n)} - \sum_{k=1}^N E_{k,c}^{(n)} \right)^{1/2} {\bar
P}_c^{(n)} P_{c,\uj}^{(n)} {\bar P}_c^{(n)} \left( {\bar
P}_c^{(n)} - \sum_{k=1}^N E_{k,c}^{(n)} \right)^{1/2}.
\end{equation} Clearly $ V_{c,\uj}^{(n)} \le {\bar P}_c^{(n)} - \sum_{k=1}^N
E_{k,c}^{(n)}$.

Put \begin{equation} V_{\uj}^{(n)} := \sum_{C \in {\cal C}_{\rm aper}} {\tilde \Pi}_C \otimes
V_{C,\uj}^{(n)} + \sum_{C' \in {\cal C}_{\rm per}} \sum_{i' \in C'} {\tilde \Pi}_{C',i'} 
\otimes V_{(C',i'),\uj}^{(n)}.
\end{equation}
This is a candidate for an additional measurement operator,
$E_{N+1}^{(n)}$, for Bob with corresponding input state $ {\tilde
\rho}_{N+1}^{(n)} = \rho_{\uj}^{(n)}= \rho_{j_1}^{(l_0)} \otimes
\rho_{j_2}^{(l_0)} \ldots \otimes \rho_{j_n}^{(l_0)}$. Clearly, the condition (i) [see below \reff{540}], is satisfied and we also have

\begin{lemma} \label{L3.32}

\begin{eqnarray} && \sum_{C \in {\cal C}_{\rm aper}} \gamma_C \tr \left[\,({\tilde
\Pi}_C \otimes V_{C,\uj}^{(n)} ) \Phi_C^{\otimes m'M + n} \left( \omega^{(m'M)}  {\bar
\rho}_{l_0}^{\otimes [n/l_0]} \right) \right]  \non \\  && \quad +
\sum_{C' \in {\cal C}_{\rm per}} \sum_{i' \in C'} \gamma_{i'} 
\tr \left[\, ({\tilde \Pi}_{C',i'} \otimes V_{(C',i),\uj}^{(n)}) 
\Phi_C^{\otimes m'M + n} \left( \omega^{(m'M)} \otimes {\bar
\rho}_{l_0}^{\otimes [n/l_0]} \right) \right] \non \\ && \qquad \qquad \qquad \leq
2^{-n[C(\Phi) - \frac{2}{3} \epsilon]}, \label{548}\end{eqnarray}
with $\gamma_{i'} = 1/L(C')$, for $i' \in C' \in {\cal{C}}_{per}$.
\end{lemma}
\textbf{Proof.} Writing ${\bar \sigma}_C^{(n)} 
= \Phi_C^{(n)}({\bar \rho}^{(n)})$, by the proof of Lemma~\ref{lem47},
the following inequality holds for an aperiodic class $C$, for $n$ large enough: 
\begin{equation} \tr ({\bar \sigma}_C^{(n)} V_{C,\uj}^{(n)})
\leq 2^{- n[{\bar{\chi}}_C - \half \epsilon]},
\label{549}
\end{equation}
where ${\bar{\chi}}_C  ={\bar{\chi}}_C ^{(l_0)}$ is given by \reff{39}, for the 
maximising ensemble, with $n=l_0$ [c.f. \reff{540}].

 Then
\begin{eqnarray} \lefteqn{\sum_{C \in {\cal C}_{\rm aper}} \gamma_C
\tr\left[ ({\tilde \Pi}_C \otimes V_{C,\uj}^{(n)})\,\Phi_C^{(m_0 M+n)} 
\left(\omega^{(m_0 M)} \otimes {\bar \rho}^{(n)} \right) \right]} \non \\ 
&\leq& \sum_{C \in {\cal C}_{\rm aper}} \gamma_C \tr\,[ {\bar\sigma}_C^{(n)}
V_{C,\uj}^{(n)} ] \non \\ &\leq & \sum_{C \in {\cal C}_{\rm aper}} \gamma_C\, 
2^{-n[{\bar{\chi}}_C - \frac{\epsilon}{2}]}
\end{eqnarray} where we used the obvious fact that ${\tilde \Pi}_C \leq \one$
and \reff{548}.

Similarly, for $i'\in C'\in {\cal{C}}_{per}$, denoting 
$Q_{n,i'} = \sum_{k=1}^N E_{k,i'}^{(n)}$, we have, using Lemma~\ref{pertyp}, 
$$ {\bar P}_{i',n}  \Phi_{C',i'}^{(n)} (\rho_{l_0}^{\otimes m}) {\bar P}_{i',n} \leq
2^{-m[S_{C'}-\frac{1}{4} \epsilon]} $$ and hence
\begin{eqnarray} \lefteqn{\tr ( {\bar\sigma}_{C',i'}^{(n)} V_{i',\uj}^{(n)})} \non \\
&=& \tr \left[ {\bar \sigma}_{C',i'}^{(n)} ({\bar P}_{C',i'}^{(n)} -
Q_{n,i'})^{1/2} {\bar P}_{C,i'}^{(n)} P_{i',\uj}^{(n)} {\bar P}_{C',i'}^{(n)}
({\bar P}_{C',i'}^{(n)} - Q_{n,i'})^{1/2} \right] \non \\
&\leq & 2^{-m[S_{C'}-\frac{1}{4} \epsilon]} 
\tr \left[ ({\bar P}_{C',i'}^{(n)} - Q_{n,i'})^{1/2} P_{i',\uj}^{(n)} 
({\bar P}_{C',i'}^{(n)} - Q_{n,i'})^{1/2} \right] \non \\
&\leq & 2^{-m[S_{C'}-\frac{1}{4} \epsilon]} 
\tr\, (P_{i',\uj}^{(n)} ) \non \\ &\leq&
2^{-n[\frac{1}{l_0} (S_{C'} - {\bar S}_{C'}) -
\frac{1}{2} \epsilon]} \nonumber\\
&\leq& 2^{-n[{\bar{\chi}}_{C'}^{(l_0)}- \frac{1}{2}]},
\label{last11}
\end{eqnarray}
where (see \reff{311})
$${\bar{\chi}}_{C'}^{(l_0)}= \frac{1}{l_0L} \sum_{i \in C'} \bigl(
S(\Phi_{C',i}^{(l_0)} ({\bar{\rho}}_{l_0})) - {\bar{S}}_i\bigr).
$$ 
In the second last inequality of \reff{last11}, we use the fact that 
$\tr\, (P_{i',\uj}^{(n)} ) \le 2^{m{\bar{S}}_{C'} + \epsilon/4}$, which is a standard
consequence of Lemma \ref{lem58}. We obtain the last line of \reff{last11} by using 
the subadditivity of the von Neumann
entropy, as in \reff{428}.

Summing $\reff{last11}$ over $i'$ and $C'$, and adding to the bound for $C \in {\cal C}_{\rm aper}$,
yields the following bound:
\bea
{\hbox{LHS of }} \reff{548} &\le & \sum_{C \in {\cal{C}}_{aper}}
\gamma_C 2^{-n[{\bar{\chi}}_C^{(l_0)} - \frac{\epsilon}{2}]}\nonumber\\
& & +  \sum_{C' \in {\cal{C}}_{per}} \sum_{i \in C'}
\gamma_{i'} 2^{-n[{\bar{\chi}}_{C'}^{(l_0)} - \frac{\epsilon}{2}]}
\eea
Now by \reff{540},
$$C(\Phi) \le 
\bigwedge_{C \in {\cal C}} {\bar{\chi}}^{(l_0)}_C + \frac{\epsilon}{6},$$
and hence
$$2^{-n[{\bar{\chi}}_{C}^{(l_0)} - \frac{\epsilon}{2}]}\le 2^{-n[C(\Phi)
- \frac{2}{3}\epsilon]},
$$
for all $C \in {\cal{C}}$, and therefore \reff{548} follows.
\qed

By maximality of ${N}$ it now follows that the condition (ii)
above cannot hold and as before we get, upon taking expectations,
\begin{corollary}
\label{corr1} \begin{eqnarray} && \sum_{C \in {\cal C}_{\rm aper}} \gamma_C\, {\mathbb{E}} \left( \tr
\left[\,({\tilde \Pi}_C \otimes V_{C,\ul{j}}^{(n)}) \Phi_C^{\otimes m_0 M + n}
\left( \omega^{(m_0 M)} \otimes {\tilde\rho}_k^{(n)} \right)  \right] \right) \non \\ 
&& + \sum_{C' \in {\cal C}_{\rm per}} \sum_{i' \in C'} \gamma_{i'}\, {\mathbb{E}} \left(\tr
\left[\,({\tilde \Pi}_{C',i'} \otimes V_{i',\ul{j}}^{(n)}) \Phi_{C',i'}^{\otimes m_0 M + n} \left( \omega^{(m_0 M)}  \otimes {\tilde\rho}_k^{(n)} \right) \right] \right) \non \\ 
&& \leq 1-\epsilon. \end{eqnarray}
\end{corollary}

\noindent We also need the following analogue of Lemma \ref{lem48}:

\begin{lemma} \label{L3.42} Assume $\eta' >  3 \delta$. Then, for $n$ large enough,
\begin{eqnarray} && \sum_{C \in {\cal C}_{\rm aper}} \gamma_C \tr
\left[\,({\tilde \Pi}_C \otimes {\bar P}_{C,n} P_{C,\ul{j}}^{(n)} {\bar P}_{C,n}) 
\Phi_C^{\otimes (m_0 M + n)}
\left( \omega^{(m_0 M)} \otimes {\rho}_{\ul{j}}^{(n)} \right)  \right] \non \\ 
&& + \sum_{C' \in {\cal C}_{\rm per}} \sum_{i' \in C'} \gamma_{i'}\, \tr
\left[\,({\tilde \Pi}_{C',i'} \otimes {\bar P}_{i',n} P_{i',\ul{j}}^{(n)} {\bar P}_{i',n}) \Phi_{C',i'}^{\otimes (m_0 M + n)}\left( \omega^{(m_0 M)}  
\otimes {\rho}_{\ul{j}}^{(n)} \right) \right] \non \\ && > 1-\eta'
\end{eqnarray} \end{lemma}
\textbf{Proof.} This is a simple consequence of Lemma~\ref{lem48} and its analogue 
for periodic classes, together with (\ref{Piaper}) and (\ref{Piper}). \qed
\begin{lemma} Assume $\eta' < \frac{1}{3} \epsilon$ and write
\begin{equation} Q_{n,C} = \sum_{k=1}^N E_{k,C}^{(n)} \quad (C \in {\cal C}_{\rm aper}) 
\mbox{ and } Q_{n,i'} = \sum_{k=1}^N E_{k,i'}^{(n)} \quad (i' \in C' \in {\cal C}_{\rm per}).
\end{equation}
Then for $n$ large enough,
\begin{eqnarray} && \sum_{C \in {\cal C}_{\rm aper}} \gamma_C \tr \left[\,({\tilde
\Pi}_C \otimes Q_{n,C} ) \Phi_C^{\otimes m'M + n} \left( \omega^{(m'M)} \otimes 
\rho_{\ul{j}}^{(n)} \right) \right]  \non \\  && \quad +
\sum_{C' \in {\cal C}_{\rm per}} \sum_{i' \in C'} \gamma_{i'} 
\tr \left[\, ({\tilde \Pi}_{C',i'} \otimes Q_{n,i'}) 
\Phi_{C',i'}^{\otimes m'M + n} \left( \omega^{(m'M)} \otimes 
\rho_{\ul{j}}^{(n)} \right) \right] \geq
(\eta')^2.\nonumber\\
\label{L3.5}\end{eqnarray} 
\end{lemma}
\textbf{Proof.} This is analogous to Lemma~\ref{lemma9}. \qed

It now follows, as before, that for $n$ large enough, ${{\tilde
N}(n)} \geq (\eta')^2 \,2^{n[C(\Phi)- \frac{2}{3} \epsilon]}.$ We
take the following states as codewords:
\begin{equation} \rho_k^{(m_0 M+n)} = \omega^{(m_0 M)} \otimes
{\tilde \rho}_k^{(n)}. \end{equation} For $n$ sufficiently large
we then have \begin{equation} N= {N_{n+m_0 M}} = {\tilde N}(n) \geq
(\eta')^2\,2^{n[C(\Phi)- \frac{2}{3} \epsilon]} \geq
2^{(m_0 M+n)[C(\Phi)- \epsilon]}.
\end{equation}

To complete the proof, we need to show that the set $\{E_k^{(n)}\}_{k=1}^N$
satisfies (\ref{inderr}). However, this follows immediately from
condition (ii) (after eq.\reff{540}):
\begin{eqnarray} \lefteqn{\tr \left[ \Phi^{(m_0 M+n)}
\left( \rho^{(m_0 M+n)}_k \right) E_k^{(n)} \right] =} \non \\ &=&
\sum_{C \in {\cal C}_{\rm aper}} \gamma_C \tr \left[\,\Phi_C^{(m_0M+n)}
\left( \omega^{(m_0 M)} \otimes
{\tilde \rho}^{(n)}_k \right) E_k^{(n)} \right] \non \\ 
&& + \sum_{C' \in {\cal C}_{\rm per}} \sum_{i \in C'} \gamma_i 
\tr \left[\,\Phi_{C',i}^{(m_0M+n)} \left( \omega^{(m_0 M)} \otimes
{\tilde \rho}^{(n)}_k \right) E_k^{(n)} \right] \non \\
&=& \sum_{C \in {\cal C}_{\rm aper}} \gamma_C \tr \left[
({\tilde \Pi}_C \otimes E_{k,C}^{(n)}) \,\Phi_C^{(m_0M+n)}
\left( \omega^{(m_0 M)} \otimes
{\tilde \rho}^{(n)}_k \right) \right] \non \\ 
&& + \sum_{C' \in {\cal C}_{\rm per}} \sum_{i \in C'} \gamma_i 
\tr \left[({\tilde \Pi}_{C',i} \otimes E_{k,i}^{(n)})\,
\Phi_{C',i}^{(m_0M+n)} \left( \omega^{(m_0 M)} \otimes
{\tilde \rho}^{(n)}_k \right) \right] \non \\ &>& 1-\epsilon.
\end{eqnarray}
\qed

\section{{Proof of the converse part of Theorem \ref{main1}}}
\label{converse}

In this section we prove that it is impossible for Alice to
transmit classical messages reliably to Bob through the channel
$\Phi$ defined by \reff{mem} and \reff{mem2} at a rate $R > C(\Phi)$.  
This is the (weak) converse part of Theorem~\ref{main1}, in the sense that the
probability of error does not tend to zero asymptotically as the
length of the code increases, for any code with rate $R >
C(\Phi)$. To prove the weak converse, suppose that Alice encodes
messages labelled by $\alpha \in {\cal M}_n$ by states
$\rho_\alpha^{(n)}$ in ${\cal B}({\cal H}^{\otimes n})$. Let the
corresponding outputs for the class $C$ of the channel be
denoted by $\sigma_{\alpha,C}^{(n)}$, i.e. \begin{equation}
\sigma_{\alpha,C}^{(n)} = \Phi_C^{(n)}(\rho_\alpha^{(n)}).
\end{equation}
Further define \begin{equation} {\bar \sigma}_C^{(n)} =
\frac{1}{|{\cal M}_n|} \sum_{\alpha \in {\cal M}_n}
\sigma_{\alpha,C}^{(n)}. \end{equation}  Let Bob's POVM
elements corresponding to the codewords $\rho_\alpha^{(n)}$ be
denoted by $E_\alpha^{(n)}$, $\alpha = 1,\dots,|{\cal M}_n|$. We
may assume that Alice's messages are produced uniformly at random
from the set ${\cal M}_n$. Then Bob's average probability of error
is given by
\begin{equation}
{\bar p}_e^{(n)} := 1- \frac{1}{|{\cal M}_n|} \sum_{\alpha \in
{\cal M}_n} \tr\,\left[ \Phi^{(n)}(\rho_\alpha^{(n)})
E_\alpha^{(n)} \right].
\end{equation} We also define the average error corresponding
to the class $C$ of the channel as
\begin{equation}
{\bar p}_{e,C}^{(n)} := 1- \frac{1}{|{\cal M}_n|} \sum_{\alpha \in
{\cal M}_n} \tr\,\left[ \Phi_C^{\otimes n}(\rho_\alpha^{(n)})
E_\alpha^{(n)} \right],
\end{equation}
so that \begin{equation} {\bar p}_e^{(n)} = \sum_{C \in {\cal C}} \gamma_C
{\bar p}_{e,C}^{(n)}. \label{errpr}
\end{equation}

Let $X^{(n)}$ be a random variable with a uniform distribution
over the set ${\cal M}_n$, characterizing the classical message
sent by Alice to Bob. Let $Y_C^{(n)}$ be the random variable
corresponding to Bob's inference of Alice's message, when the
codeword is transmitted through the class $C$. 
It is defined by the conditional probabilities
\begin{equation}
{\mathbb{P}}\,[{Y_C^{(n)}} = \beta\,|\, X^{(n)} = \alpha] = \tr\,
[\Phi_C^{(n)}(\rho_{\alpha}^{(n)}) E_{\beta}^{(n)}].
\end{equation} By Fano's inequality,
\begin{equation} h({\bar p}_{e, C}^{(n)}) + {\bar p}_{C,e}^{(n)}
\log(|{\cal M}_n|-1) \geq H(X^{(n)}\,|\, Y_C^{(n)}) = H(X^{(n)}) -
H(X^{(n)}\,:\, Y_C^{(n)}). \label{Fano}
\end{equation}
Here $h(\cdot)$ denotes the binary entropy and $H(\cdot)$ denotes
the Shannon entropy. By the Holevo bound, for $C \in {\cal{C}}_{aper}$ we have
\begin{eqnarray}
\lefteqn{H(X^{(n)}\,:Y_C^{(n)})} \non \\ &\leq& 
S\left(\frac{1}{|{\cal M}_n|}
\sum_{\alpha \in {\cal M}_n} \Phi_C^{(n)}
(\rho_{\alpha}^{(n)}) \right) - \frac{1}{|{\cal M}_n|}
\sum_{\alpha \in {\cal M}_n} S \left( \Phi_C^{(n)}
(\rho_{\alpha}^{(n)}) \right) \nonumber  \\
&=&  n{\bar\chi}_C \left( \left\{ \frac{1}{|{\cal M}_n|},
\rho_{\alpha}^{(n)} \right\}_{\alpha \in {\cal M}_n} \right), 
\label{Holevobnd}
\end{eqnarray}
where ${\displaystyle{{\bar\chi}_C \left( \left\{ \frac{1}{|{\cal M}_n|},
\rho_{\alpha}^{(n)} \right\}_{\alpha \in {\cal M}_n} \right)}}$ is given by
\reff{311}. 

For $C \in {\cal{C}}_{per}$, with period $L$,
\bea
\lefteqn{H(X^{(n)}\,:Y_C^{(n)})} \non \\ &\leq& 
S\left(\frac{1}{|{\cal M}_n|}
\sum_{\alpha \in {\cal M}_n}\frac{1}{L}\sum_{i\in C} \Phi_{C,i}^{(n)}
(\rho_{\alpha}^{(n)}) \right) - \frac{1}{|{\cal M}_n|}
\sum_{\alpha \in {\cal M}_n} S \left( \frac{1}{L}\sum_{i\in C}\Phi_{C,i}^{(n)}
(\rho_{\alpha}^{(n)}) \right) \nonumber  \\
&=& \frac{1}{|{\cal M}_n|} \sum_{\alpha \in {\cal M}_n} S\bigl(  \frac{1}{L}\sum_{i\in C}\Phi_{C,i}^{(n)}
(\rho_{\alpha}^{(n)}) ||\frac{1}{|{\cal M}_n|} \sum_{\beta \in {\cal M}_n} \frac{1}{L}\sum_{i\in C}\Phi_{C,i}^{(n)}
(\rho_{\beta}^{(n)}) \bigr)\nonumber\\
&\le & \frac{1}{|{\cal M}_n|L}\sum_{\alpha \in {\cal M}_n}\sum_{i\in C}S\bigl( \Phi_{C,i}^{(n)}
(\rho_{\alpha}^{(n)})|| \frac{1}{|{\cal M}_n|}\sum_{\beta \in {\cal M}_n}\Phi_{C,i}^{(n)}
(\rho_{\beta}^{(n)})\bigr)\nonumber\\
&=& \frac{1}{L} \sum_{i\in C} \chi_{C,i}^{(n)} \left( \{\frac{1}{|{\cal M}_n|},\rho_{\alpha}^{(n)}\} \right)
\nonumber\\
&=& n{\bar{\chi}}_{C}^{(n)} \left( \{\frac{1}{|{\cal M}_n|},\rho_{\alpha}^{(n)}\} \right).
\eea
In the above, we use the convexity of the relative entropy $S(\sigma||\omega) := \tr \sigma (\log \sigma - \log \omega)$,
for density matrices $\sigma$ and $\omega$. 

Therefore, for any class $C$ we have the upper bound
\be H(X^{(n)}\,:Y_C^{(n)}) \le n{\bar{\chi}}_{C}^{(n)} \left( \{\frac{1}{|{\cal M}_n|},\rho_{\alpha}^{(n)}\} \right).
\ee
Inserting this into Fano's inequality, \reff{Fano}, now yields 
\begin{equation} h({\bar p}_{C,e}^{(n)}) + {\bar p}_{C,e}^{(n)} \log\,{|{\cal M}_n|} \geq
\log\,{|{\cal M}_n|} - n\,{\bar{\chi}}_C \left( \left\{ \frac{1}{|{\cal
M}_n|} , \rho_{\alpha}^{(n)} \right\}_{\alpha} \right).
\end{equation} However, since
\begin{equation} C(\Phi) \geq \bigwedge_{C \in {\cal C}} 
{\bar{\chi}}_C \left( \left\{ \frac{1}{ |{\cal M}_n|} , \rho_{\alpha}^{(n)}
\right\}_{\alpha} \right)
\end{equation} and $R = \frac{1}{n} \log |{\cal M}_n| > C(\Phi)$,
there must be at least one class $C$ such that
\begin{equation} {\bar p}_{e,C}^{(n)} \geq 1 - \frac{C(\Phi)+
{1}/{n}}{R} > 0.\label{errpr2} \end{equation} We conclude from
(\ref{errpr}) and \reff{errpr2} that \begin{equation} {\bar
p}_e^{(n)} \geq  \left(1 - \frac{C(\Phi)+ {1}/{n}}{R}
\right)\,\bigwedge_{C \in {\cal C}} {\gamma_C}.
\end{equation} \qed

\subsection*{Remark} Note that the strong converse property \cite{hayashi, winter} does not hold for general Markovian
channels. For example, for a convex combination of memoryless channels\footnote{A
classical version of such a channel was introduced by
Jacobs \cite{jacobs} and studied further by
Ahlswede \cite{ahlswede}, who obtained an expression for
its capacity.}:
\begin{equation} \Phi^{(n)}(\rho^{(n)}) = \sum_{i=1}^M
\gamma_i \Phi_i^{\otimes n}(\rho^{(n)}),
\label{def_channel}
\end{equation}
where
$\Phi_i: {\cal B}({\cal H}) \to {\cal B}({\cal })$, Bob's error probability does {\em{not}}
tend to $1$ asymptotically in $n$ for a rate $R$, such that $C(\Phi) < R <{\bar{C}}(\Phi)$,
where
           $${\bar{C}}(\Phi) := \bigvee_{i=1}^M \chi_i^*,$$
and $\chi_i^*$ denotes the Holevo capacity \cite{holevo, SW} of the memoryless channel $\Phi_i$.

\section*{Acknowledgements} TCD would like to acknowledge the hospitality
of the Statistical Laboratory of Cambridge University during his sabbatical. 
The work was supported by the European Commission
through the Integrated Projects SECOQC and FET/QIPC "SCALA".

\section*{Appendix A}
\begin{lemma} \label{lema1} If $\Phi^{(n)}$ is a quantum channel with memory of the form 
(\ref{mem}). Then the limit in (\ref{314}) exists. In particular, the limit in \reff{cap1} exists.
\end{lemma}
\textbf{Proof.} Denote \begin{equation} {\bar{\chi}}_n = \sup_{\{p_j^{(n)},\rho_j^{(n)}\}}\bigwedge_{C\in {\cal{C}}} 
{\bar{\chi}}_C^{(n)}(\{p_j^{(n)},\Phi^{(n)}
(\rho_j^{(n)})\}).\label{574} \end{equation} We shall prove that for any $\delta > 0$ there exist
$n_0$ and $m_0$ such that for all $n' \geq n_0$ and $n \geq m_0 n'$, ${\bar{\chi}}_n \geq {\bar{\chi}}_{n'} - \delta$. This proves the lemma because obviously, $0 \leq {\bar{\chi}}_n \leq \log {\hbox{ dim}} {\cal{K}}$, and it follows that 
$$ \liminf_{n \rightarrow \infty} {\bar{\chi}}_n \geq {\bar{\chi}}_{n'} - \delta $$ and hence ${\displaystyle{\liminf_{n \rightarrow \infty} {\bar{\chi}}_n \geq \limsup_{n' \rightarrow \infty} {\bar{\chi}}_{n'} - \delta}}$ where $\delta > 0$ is arbitrary. 

To prove the statement, let $n'$ be large, and suppose that $\{p_j^{(n')}, \rho_j^{(n')} \}$ is a maximising ensemble
for \reff{574}, with $n$ replaced by $n'$. Given $n \geq n'$, put $m = [n/n']$ and $l = n-mn'$. Define the states $\rho^{(n)}_{\ul{j}} = \bigotimes_{r=1}^m \rho_{j_r}^{(n')} \otimes \rho_{j_{m+1}}^{(l)}, $ where $\rho_j^{(l)}$ is the reduced state on ${\cal H}^{\otimes l}$. 
Then ${\bar \rho}^{(n)} = \otimes_{r=1}^m {\bar \rho}^{(n')} \otimes {\bar \rho}^{(l)}$,
with ${\bar \rho}^{(n')}:= \sum_j p_j^{(n')} \rho_j^{(n')}$. We now write for any 
class $C \in {\cal{C}}$, \begin{eqnarray} \Phi_C^{(n)}({\bar \rho}^{(n)}) &=&
\sum_{i_1,\dots,i_{m+1} \in C} \sum_{i'_1,\dots,i'_{m+1} \in C}
\frac{q_{i'_1i_2}}{\gamma_{i_2}} \dots
\frac{q_{i'_{m}i_{m+1}}}{\gamma_{i_{m+1}}} \non \\ && \qquad \times
{\sigma}_C^{(n')}(i_1,i'_1) \otimes \dots \otimes {\sigma}_C^{(n')}(i_m,i'_m) 
\otimes \sigma_C^{(l)} (i_{m+1},i'_{m+1}), \non \\ && \end{eqnarray} where 
\begin{eqnarray} {\sigma}_C^{(n')}(i,i') &=& \sum_{i_2,\dots,i_{n'-1} \in
C} \gamma_i q_{ii_2} q_{i_2i_3} \dots q_{i_{n'-1}i'} \non \\
&& \qquad \times (\Phi_i \otimes \Phi_{i_2} \otimes \dots \otimes
\Phi_{i'})({\bar \rho}^{(n')}) \end{eqnarray} and similarly for $\sigma_C^{(l)}(i,i')$. 
Let $\gamma = {\bigwedge}_{i \in I} \gamma_i$. Using positivity of the density
operators and the fact that $q_{ij} \leq
1 \leq \gamma_i/\gamma$, we obtain the simple operator inequality
\begin{equation} \Phi_C^{(n)}({\bar \rho}^{(n)}) \leq
\frac{1}{\gamma^{m}} \Phi_C^{(n')}({\bar \rho}^{(n')}) \otimes \dots \otimes 
\Phi_C^{(n')}({\bar \rho}^{(n')}) \otimes \Phi_C^{(l)}({\bar \rho}^{(l)}). \end{equation} 
Inserting this into the definition of $S(\Phi^{(n)}({\bar \rho}^{(n)}))$ and
using the operator monotonicity of the logarithm and the fact that
$(\gamma_i)$ is the equilibrium distribution, i.e. $\sum_{i \in I}
\gamma_i q_{ij} = \gamma_j$, we obtain \begin{equation}
S(\Phi_C^{(n)}({\rho}^{(n)})) \geq m S(\Phi_C^{(n')}({\bar \rho}^{(n')})) 
+ S(\Phi_C^{(l)}({\bar \rho}^{(l)})) + m \log \gamma. \end{equation}
On the other hand, by subadditivity, \begin{equation} 
S(\Phi_C^{(n)}(\rho^{(n)}_{\ul{j}})) \leq 
\sum_{r=1}^m  S(\Phi_C^{(n')}(\rho^{(n')}_{j_r})) + S(\Phi_C^{(l)}(\rho_{j_{m+1}}^{(l)})) 
\end{equation} so that 
\begin{equation} {\bar{\chi}}_C^{(n)} \left( \{ p_{\ul{j}}^{(n)},\Phi^{(n)}(\rho_{\ul{j}}^{(n)}) \} \right) \geq \frac{m n'}{n} 
{\bar{\chi}}_{n'} + \frac{m}{n} \log \gamma,
 \end{equation} 
for all $C \in {\cal{C}}$.
\qed

\section*{Appendix B}
{\bf{Proof of Lemma \ref{lem44}:}} Let $l_1$ be so large that \begin{equation}
S_M({\bar \phi}_\infty) \leq \frac{1}{l_1} S({\bar \sigma}^{(l_1)}) 
< S_M({\bar \phi}_\infty) + \frac{\epsilon}{8}.
\end{equation} Let $\Omega = \{\lambda_k\}$ denote the spectrum
of ${\bar\sigma}^{(l_1)}$, and let $\pi_k$ be the projection onto
the eigenvector with eigenvalue $\lambda_k$. For any $r>0$ and $C \subset {\cal
X}^r$, put \begin{equation} q_C = \sum_{(\lambda_{k_1}, \dots,
\lambda_{k_r}) \in C} \pi_{k_1} \otimes \dots \otimes \pi_{k_r},
\end{equation} and define the probability measures $\nu_r$ on $\Omega^r$ and
$\nu_\infty$ on $\Omega^{{\mathbb{N}}}$ by \begin{equation} \nu_r(C) =
\Tr(\Phi^{(rl_0l_1)}({\bar \rho}_{l_0}^{\otimes (rl_1)}) q_C)
\mbox{ and } \nu_\infty (C) = {\bar \phi}_\infty (q_C).
\end{equation} By Lemma~\ref{lemma2}, $\nu_\infty$ is ergodic and by
McMillan's theorem \cite{McMillan} there exists a typical set
\begin{eqnarray} T_\epsilon^{(r)} &=& \left\{ (\lambda_{k_1},
\dots,\lambda_{k_r}) \in \Omega^r\,| \right. \non \\ && \qquad
\left. 2^{-r (h_{KS}(\nu_\infty) + \epsilon/{8})} \leq \nu_r
(\{(\lambda_{k_1},\dots,\lambda_{k_r})\}) \leq
2^{-r(h_{KS}(\nu_\infty) -\epsilon/{8})} \right\}, \non \\ &&
\label{583}
\end{eqnarray} satisfying
\begin{equation} \nu_r (T_\epsilon^{(r)}) > 1-\delta^2
\end{equation} for $r$ large enough, where $h_{KS}(\nu_\infty)$ 
denotes the Kolmogorov-Sinai entropy.
Now, \begin{equation} h_{KS}(\nu_\infty) = \inf_r \frac{1}{r}
H(\nu_r) \leq H(\nu_1) = S({\bar\sigma}^{(l_1)}) < l_1 \left(S_M ({\bar\phi}_\infty) +
\frac{\epsilon}{8} \right),
\label{585}
\end{equation} 
where $H(\nu)$ denotes the Shannon entropy corresponding to
the probability measure $\nu$.
On the other hand \begin{equation}
h_{KS}(\nu_\infty) \geq  l_1\,S_M ({\bar\phi}_\infty)
\label{586}
\end{equation} because, by positivity of the relative entropy,
\begin{eqnarray} \lefteqn{S ({\bar \sigma}^{(rl_1)})} \non \\ 
&=& - \Tr \left[ {\bar\sigma}^{(rl_1)} 
\log {\bar\sigma}^{(rl_1)} \right] \non \\
&\leq & -\Tr \left[ {\bar\sigma}^{(rl_1)} \log \left(
\opplus\limits_{k_1,\dots,k_r} \bigl[\Tr \left( {\bar\sigma}^{(rl_1)}\bigr]
(\pi_{k_1} \otimes \dots \otimes \pi_{k_r})
\right) \pi_{k_1} \otimes \dots \otimes \pi_{k_r} \right) \right] \non \\
&=& - \sum_{k_1,\dots,k_r}  \Tr [{\bar\sigma}^{(rl_1)} \pi_{k_1}
\otimes \dots \otimes \pi_{k_r}] %\non \\ && \quad  \times
\log \Tr [{\bar\sigma}^{(rl_1)} \pi_{k_1} \otimes \dots \otimes
\pi_{k_r}] \non \\ &=& H(\nu_r).
\end{eqnarray} 
For arbitrary $m$, let $r = [m/l_1]$ and define
\begin{equation} \pi_{\ul{k}}^{(m)} = \pi_{k_1} \otimes \dots
\otimes \pi_{k_r} \otimes \one \in {\cal{B}}({\cal{K}}_{l_0}^{\otimes m}), 
\quad \ul{k} = (k_1,\dots,k_r). 
\end{equation} Let
$\ol{T}_\epsilon^{(m)} = \{\ul{k}:\, (\lambda_{k_1},
\dots,\lambda_{k_r}) \in T_\epsilon^{(r)} \}$, and define 
$${\cal T}_\epsilon^{(m)} =\oplus_{{\underline{k}} \in \ol{T}_\epsilon^{(m)}}
\pi_{\ul{k}}^{(m)}({\cal K}_{l_0}^{\otimes m})$$. Clearly,
\begin{equation} {\bar\phi}_\infty \left(
\oplus_{\ul{k} \in \ol{T}_\epsilon^{(m)}} \pi_{\ul{k}}^{(m)} \right) =
\Tr [{\bar\sigma}^{(rl_1)} q_{{T}_\epsilon^{(r)}}] = 
\nu_r \left( T_\epsilon^{(r)} \right) > 1-\delta^2.
\end{equation} Moreover, if $\ul{k} \in \ol{T}_\epsilon^{(m)}$, it
follows from \reff{583}, \reff{585} and \reff{586}
that
\begin{equation}  \frac{1}{m}
\log \nu_r(\{(\lambda_{k_1}, \dots,\lambda_{k_r})\}) \leq -
\frac{rl_1}{m} \left(S_M({\bar\phi}_\infty) - 
\frac{1}{l_1} \frac{\epsilon}{8} \right),
\end{equation} and
\begin{equation} \frac{1}{m}
\log \nu_r(\{(\lambda_{k_1}, \dots,\lambda_{k_r})\}) \geq -
\frac{rl_1}{m} \left(S_M({\bar\phi}_\infty) + \left(1 + \frac{1}{l_1} \right) 
\frac{\epsilon}{8} \right).
\end{equation} Taking $l_1 > 3$ and $m$ large enough, we obtain
\begin{equation} \left| \frac{1}{m} \log  \nu_r(\{(\lambda_{k_1}, \dots,\lambda_{k_r})\}) +
S_M({\bar\phi}_\infty) \right| < \frac{\epsilon}{6}. \end{equation}

Now let \begin{equation} {\bar P}_{ml_0} = \bigoplus_{\ul{k} \in
\ol{T}_\epsilon^{(m)}} \pi_{\ul{k}}^{(m)} \end{equation} and assume
that $l_1$ is so large that ${\epsilon l_1/12} > - \log \gamma_{\rm
min}$, where $\gamma_{\rm min} = \bigwedge_{i\in I} \gamma_i$.
Note that $\gamma_{\rm min} > 0$. Define
\begin{equation} {\bar \sigma}_l(i,i') = \sum_{j_1,\dots,j_l=1}^J
p_{\ul{j}}^{(l)} \sum_{i_2, \dots, i_{l-1}} \gamma_{i} q_{ii_2}
\dots q_{i_{l-1}i'} \Phi_{i}(\rho_{j_1}) \otimes \dots \otimes
\Phi_{i'} (\rho_{j_l}),
\end{equation}
where $\ul{j} = (j_1, j_2, \ldots, j_l)$.
Then we can write as in the proof of Lemma \ref{lemma1},
\begin{eqnarray} \label{sigmaexpr}
{\bar\sigma}_{ml_0} &=& \sum_{i_1,\dots,i_{2r+2}}
\frac{q_{i_2i_3}}{\gamma_{i_3}} \,
\frac{q_{i_4i_5}}{\gamma_{i_5}} \dots
\frac{q_{i_{2r}i_{2r+1}}}{\gamma_{i_{2r+1}}} \non \\ && \qquad
\times  {\bar\sigma}_{l_1}(i_1,i_2) \otimes \dots \otimes
{\bar\sigma}_{l_1}(i_{2r-1},i_{2r}) \otimes {\bar\sigma}^{(m-rl_1)}
(i_{2r+1},i_{2r+2}). \non \\ && \end{eqnarray} 
Using the positivity of
the transition probabilities, we have 
$${\bar P}_{ml_0} {\bar\sigma}_{ml_0} {\bar P}_{ml_0}  \le 2^{-m[S_M({\bar{\phi}}_\infty)
- \frac{\epsilon}{4}]} {\one}^{(ml_0)}.$$

By the fact that $\pi_k$ is an
eigenprojection of ${\bar\sigma}^{(l_1)}$ we then have
\begin{equation} {\bar P}_{ml_0} {\bar\sigma}_{ml_0} {\bar P}_{ml_0} 
\leq  \gamma_{\rm min}^{-r} 2^{-m
(S_M({\bar\phi}_\infty)-\epsilon/{6})} \one^{(ml_0)} . \end{equation} But
$\gamma_{\rm min}^{-r} < 2^{-m \epsilon/{12}}$ by the above
assumption. \qed

\section*{Appendix C}
{\bf{Proof of Lemma \ref{lem46}}} In the following, we suppress the dependence on $l_0$.
We follow Hiai \& Petz \cite{hiai}, as in
Lemma~\ref{lem44}. Fix $l \geq 12$ large enough so that
\begin{equation} \frac{1}{l} S(\Sigma_{ll_0})< {S}_M(\psi_\infty) -
\frac{\epsilon}{12}.
\label{422}
\end{equation} 
Let ${\cal Y}_{\ul{j}}^{(l)}$ be the
spectrum of  $\sigma_{\ul{j}}^{(ll_0)} =
\Phi^{(ll_0)}(\rho_{j_1}^{(l_0)}\otimes \rho_{j_2}^{(l_0)}\ldots \otimes \rho_{j_l}^{(l_0)})$. 
Note that $\Sigma_{ll_0}$ can be
represented as a block-diagonal matrix in
$\bigoplus_{j_1,\dots,j_l = 1}^J {\cal K}_{l_0}^{\otimes l}$ with
spectrum consisting of eigenvalues $\nu_{\ul{j},k} =
p_{\ul{j}}^{(l)} \alpha_{\ul{j},k}$ with $\ul{j} \in
\{1,\dots,J\}^l$, $k = 1,\dots,(\mbox{dim}\,({\cal K}_{l_0}))^l$,
and $\alpha_{\ul{j},k}$ being the eigenvalues of $\sigma_{\ul{j}}^{(ll_0)}$. Let
\begin{equation} {\cal Y}_l = \bigcup_{\ul{j} \in \{1,\dots,J\}^l}
{\cal Y}_{\ul{j}}^{(l)}. \end{equation}  We now define measures
$\mu_s$, for $s \in \mathbb{N}$, on $\left({\cal Y}_l\right)^s$ by
\begin{equation} \mu_s(C) = \sum_{\ul{j} \in \{1,\dots,J\}^{sl}}
p_{\ul{j}}^{(sl)} \Tr \left( \sigma_{\ul{j}}^{(sl)} q_{C}^{(s)}
\right),
\end{equation} where
$C\subset ({{\cal Y}_l})^s$, and 
\begin{equation} q_{C}^{(s)} =
\sum_{(\lambda_{\ul{j}_1,k_1},\dots,\lambda_{\ul{j}_s,k_s}) \in C}
\pi_{\ul{j}_1,k_1} \otimes \dots \otimes \pi_{\ul{j}_s,k_s}
\end{equation} for $\ul{j} = (\ul{j}_1,\dots,\ul{j}_s)$.
(Here $\pi_{\ul{j},k}$ denotes the projection onto the $k$-th
eigenvector of $\sigma_{\ul{j}}^{(l)}$.) We also define the
projective limit $\mu_\infty$ on ${\cal Y}_l^{{\mathbb{N}}}$ by
\begin{equation} \mu_\infty (C) = \mu_s (C) = \psi_\infty
(q_C^{(s)}), \end{equation} for a cylinder set $C \in ({\cal
Y}_l)^s$. It follows from Lemma~\ref{lem45}  that $\mu_\infty$ is
ergodic. Define typical sets \begin{eqnarray} {\tilde
T}_{\ul{j},\epsilon}^{(s)} &=& \left\{ (\lambda_{\ul{j}_1,k_1},
\dots,\lambda_{\ul{j}_s,k_s}) \in {\cal Y}_l^s\,| \right. \non \\
&&  \left. 2^{-s (h_{KS}(\mu_\infty) + \epsilon/12)} \leq \mu_s
(\{(\lambda_{\ul{j}_1,k_1},\dots,\lambda_{\ul{j}_s,k_s})\}) \leq
2^{-s(h_{KS}(\mu_\infty) -\epsilon/12)} \right\}, \non \\ &&
\end{eqnarray} where $h_{KS}(\mu_\infty)$ is the
Kolmogorov-Sinai entropy of $\mu_\infty$. By McMillan's theorem \cite{McMillan},
\begin{equation} \mu_s \left( \bigcup_{\ul{j}}
{\tilde T}_{\ul{j},\epsilon}^{(s)} \right) > 1-\half\delta^2
\end{equation} for $s$ large enough.
Now, \begin{eqnarray} h_{KS}(\mu_\infty) &=& \inf_s \frac{1}{s}
H(\mu_s) \non \\ &\leq& H(\mu_1) = S(\Sigma_l) \non \\ 
&<& l \left({S}_M(\psi_\infty) + \frac{\epsilon}{12} \right), 
\end{eqnarray} 
by \reff{422}, 
and on the other hand \begin{equation}
h_{KS}(\mu_\infty) \geq  l S_M(\psi_\infty)
\end{equation} by positivity of the relative entropy.

For arbitrary $m$ we argue as in Lemma~\ref{lem24}, and let $s =
[m/l]$. Writing, $m = sl+r$, and $\ul{j} = (j_1,\dots,j_m) =
(\ul{j}_1,\dots,\ul{j}_s, \ul{j}_0)$, we have
\begin{equation} \pi_{\ul{j},\ul{k}}^{(ml_0)} = \pi_{\ul{j}_1,k_1}
\otimes \dots \otimes \pi_{\ul{j}_s,k_s} \otimes
\pi_{\ul{j}_0}^{(r)},
\end{equation} where $\pi_{\ul{j}_0}^{(r)}$ is the projection in
$\bigoplus_{j_1,\dots,j_r=1}^J {\cal K}^{\otimes r}$ onto the
$\ul{j}_0$-th summand.  Let ${\tilde T}_{\ul{j},\epsilon}^{[m]} =
 {\tilde T}_{\ul{j},\epsilon}^{(s)} $. 
Then,
\begin{eqnarray} \psi_\infty \left(
\bigoplus_{\ul{j} \in \{1,\dots,J\}^m} \bigoplus_{\ul{k} \in
{\tilde T}_{\ul{j},\epsilon}^{(m)}} \pi_{\ul{j},\ul{k}}^{(ml_0)} \right) &=& \Tr
\left[ \Sigma_{sl} \left(\bigoplus\limits_{\ul{j} \in
\{1,\dots,J\}^{sl}} q_{{\tilde T}_{\ul{j},\epsilon}^{(s)}} \right)
\right] \non \\ &=& \mu_s \left( \bigcup_{\ul{j}} {\tilde
T}_{\ul{j},\epsilon}^{(s)} \right) > 1-\half\delta^2.
\end{eqnarray} Moreover, if $(\lambda_{\ul{j}_1,k_1},
\dots,\lambda_{\ul{j}_s,k_s}) \in {\tilde T}_{\ul{j},\epsilon}^{[m]}$,
\begin{equation}
\frac{1}{m} \log \mu_s\left(\{(\lambda_{\ul{j}_1,k_1},
\dots,\lambda_{\ul{j}_s,k_s})\} \right) \leq - \frac{sl}{m} \left(
{S}_M(\psi_\infty)- \frac{1}{l} \frac{\epsilon}{12} \right),
\end{equation} and
\begin{equation} \frac{1}{m} \log \mu_s \left( \{(\lambda_{\ul{j}_1,k_1},
\dots,\lambda_{\ul{j}_s,k_s})\} \right) \geq - \frac{sl}{m} \left(
{S}_M(\psi_\infty) + \left(1+\frac{1}{l}\right) \frac{\epsilon}{12} \epsilon
\right).
\end{equation}

Finally define the typical set of indices $\ul{j}$:
\begin{equation} T_\epsilon^{[m]} = \left\{ \ul{j} \in
\{1,\dots,J\}^m\,|\, 2^{-m (H(\{p_j\})+\epsilon/12)} \leq
p_{\ul{j}}^{(m)} \leq 2^{-m (H(\{p_j\})-\epsilon/12)} \right\}.
\end{equation} Then for $m$ large enough, \begin{equation}
\mathbb{P}^{\otimes m} \left[ T_\epsilon^{[m]} \right] > 1 - \half \delta^2,
\end{equation} if $\mathbb{P}$ denotes the probability with respect
to the ensemble probabilities $\{p_j\}_{j=1}^J$. Defining
\begin{equation} T_{\ul{j},\epsilon}^{(m)} = \left\{ \begin{array}{ll}
T_{\ul{j},\epsilon}^{[m]} &\mbox{ if } \ul{j} \in T_\epsilon^{[m]} \\
\emptyset &\mbox{ if } \ul{j} \notin T_\epsilon^{[m]}, \end{array}
\right. \end{equation} we have for $(\lambda_{\ul{j}_1,k_1},
\dots,\lambda_{\ul{j}_s,k_s}) \in T_{\ul{j},\epsilon}^{(m)}$,
\begin{eqnarray} \frac{1}{m} \log
\lambda_{\ul{j},\ul{k}}^{m)} &=& -\frac{1}{m} \log (p_{\ul{j}_1}^{(l)}
\dots p_{\ul{j}_s}^{(l)}) + \frac{1}{m} \log
\mu_s(\{(\lambda_{\ul{j}_1,k_1}, \dots,\lambda_{\ul{j}_s,k_s})\})
\non \\ &\leq & - \frac{sl}{m} \left( {S}_M (\psi_\infty)-
\frac{1}{l} \frac{\epsilon}{12} \right) + H(\{p_j\}) + \frac{1}{12} \epsilon
\non \\ &\leq & - {\bar S}_M + \frac{1}{4} \epsilon,
\end{eqnarray} and
\begin{eqnarray} \frac{1}{m} \log
\lambda_{\ul{j},\ul{k}}^{m)} &=& -\frac{1}{m} \log (p_{\ul{j}_1}^{(l)}
\dots p_{\ul{j}_s}^{(l)}) + \frac{1}{m} \log
\mu_s(\{(\lambda_{\ul{j}_1,k_1}, \dots,\lambda_{\ul{j}_s,k_s})\})
\non \\ &\geq& - \left( {\bar S}_M  + \frac{1}{4} \epsilon \right)
\end{eqnarray} for $m$ large enough. 
Moreover, \begin{eqnarray} \psi_\infty \left(
\bigoplus_{\ul{j} \in T_\epsilon^{[m]}} \bigoplus_{\ul{k} \in
T_{\ul{j},\epsilon}^{(m)}} \pi_{\ul{j},\ul{k}}^{(ml_0)} \right) &=& \Tr
\left[ \Sigma_{m} \left(\opplus\limits_{\ul{j} \in T_\epsilon^{[m]}}
q_{{\tilde T}_{\ul{j},\epsilon}^{(s)}} \right) \right] \non \\ &=&
\sum_{\ul{j} \in T_\epsilon^{[m]}} p_{\ul{j}}^{(m)} \Tr \left(
\Phi^{(m)}(\rho_{\ul{j}}^{(m)}) q_{{\tilde T}_{\ul{j},\epsilon}^{(s)}}
\right) \non \\ &\geq&  \mu_s \left( \bigcup_{\ul{j}} {\tilde
T}_{\ul{j},\epsilon}^{(s)} \right) - \mathbb{P}^{\otimes m} \left[ \left(
T_\epsilon^{[m]} \right)^c \right] \non \\ &>& 1-\delta^2.
\end{eqnarray}
 \qed

\end{document}